\documentclass[a4paper,oneside,12pt,numbers=noenddot]{scrartcl}
\usepackage[T1]{fontenc}
\usepackage[ansinew]{inputenc}
\usepackage{lmodern}
\usepackage[a4paper,includehead,includefoot]{geometry}
\usepackage{natbib}
\usepackage{hyperref}
\usepackage{amsmath}
\usepackage{amsthm}
\usepackage{amssymb}
\usepackage{amsfonts}
\usepackage{bbm}
\usepackage{graphicx}
\allowdisplaybreaks
\usepackage{setspace}
\begin{document}

\noindent
\begin{center}
\sffamily{\textbf{\LARGE{Incorporating delayed entry into the joint frailty model for recurrent events and a terminal event }}}
\end{center}
\vskip0.1cm
\renewcommand*{\thefootnote}{\fnsymbol{footnote}}
\noindent\normalfont
\large{\textbf{Marie B\"ohnstedt\footnote{Address for correspondence: \sffamily{e-mail: boehnstedt.marie@gmail.com}\vspace{0.2cm}}{\let\thefootnote\relax\footnote{\hspace{-0.25cm}This preprint  was last edited March 26, 2021, and has not undergone peer review or any post-submission improvements or corrections. The Version of Record of this article is published in \textit{Lifetime Data Analysis}, and is available online at https://doi.org/10.1007/s10985-022-09587-z.}}$^{,1,2}$, Jutta Gampe$^1$, Monique A.~A.~Caljouw$^3$, and Hein Putter$^2$}}
\vskip0.2cm\normalsize\noindent
$^1$ Max Planck Institute for Demographic Research,  Rostock, Germany\\
$^2$ Department of Biomedical Data Sciences, Leiden University Medical Center,\\ Leiden, The Netherlands\\
$^3$ Department of Public Health and Primary Care, Leiden University Medical Center, Leiden, The Netherlands
\vskip0.5cm
\normalfont
\noindent
\textbf{Abstract:} In studies of recurrent events, joint modeling approaches are often needed to allow for potential dependent censoring by a terminal event such as death. Joint frailty models for recurrent events and death with an additional dependence parameter have been studied for cases in which individuals are observed from the start of the event processes. However, the samples are often selected at a later time, which results in delayed entry. Thus, only individuals who have not yet experienced the terminal event will be included in the study. We propose a method for estimating the joint frailty model from such left-truncated data. The frailty distribution among the selected survivors differs from the frailty distribution in the underlying population if the recurrence process and the terminal event are associated. The correctly adjusted marginal likelihood can be expressed as a ratio of two integrals over the frailty distribution, which may be approximated using Gaussian quadrature. The baseline rates are specified as piecewise constant functions, and the covariates are assumed to have multiplicative effects on the event rates.
We assess the performance of the estimation procedure in a simulation study, and apply the method to estimate age-specific rates of recurrent urinary tract infections and mortality in an older population.
\normalfont\normalsize

\section{Introduction}

Repeated occurrences of the same type of event in one individual arise in various applications. Examples of such recurrent event data include incidents of myocardial infarction, recurrent infections, fractures, or tumor relapses.

If the individual is additionally at risk of experiencing a terminal event such as death, which will stop the recurrent event process, this might induce dependent censoring of the recurrence process. Therefore, approaches for jointly modeling the two processes of the recurrent events and the terminal event have been developed. Moreover, studies might explicitly address the question of whether there is an association between the processes by asking, for instance, whether individuals who experience more recurrences also have a higher risk of experiencing the terminal event. Specific joint models can provide additional insights into the direction and the strength of the association between the event processes.

The choice of the time scale~$t$, along which the recurrence process and the terminal event process are assumed to evolve, depends on the specific application. In clinical studies, the time since randomization is often used as the time scale; whereas in demographic studies of fertility, for instance, the most relevant time scale is age.  

For a given time scale, the event processes can, in some cases, be observed from the time origin~$t_0$. Consider, for example, a medical study in which the time since the disease onset or diagnosis is used as the time scale. If each individual is observed from the disease onset or diagnosis onwards, then all individuals enter the study at time~$t_0=0$.

However, studies are often initiated at a later point in time when the two processes have already started. If in the above setting patients do not participate in the clinical study until some period of time after their diagnosis, individual times of study entry will differ from~0, and might also vary between patients. Similarly, in studies of certain diseases in old-age populations, such as in register-based studies of cardiovascular disease and mortality, age can be considered the natural time scale \citep[see, for instance,][]{Modigetal:2013}. The individuals included in such analyses have already reached a certain advanced age by the beginning of the study period.

In these cases, the data are left-truncated in the sense that the individuals can enter the study only if they have not yet experienced the terminal event. If the recurrence process and the terminal event process are associated, this sample of survivors is not a random sample of the underlying population. Rather, the sample is comprised of individuals who tend to have a lower risk of experiencing the terminal event, and -- if there is a positive association between the two processes -- to also have a lower risk of experiencing recurrent events.
Thus, to obtain valid inferences, correctly adjusting for the truncation is crucial.

The example we use in this study to illustrate our method focuses on recurrent urinary tract infections (UTIs) in older residents of long-term care facilities (LTCF). The original study by \citet{Caljouw:2014} investigated whether cranberry capsules are effective in preventing UTIs. Given that around 34\% of the elderly study population died during the follow-up period, recurrent UTIs and mortality had to be modeled jointly. The original study used time since randomization as the time scale. By contrast, we have chosen to use age as the main time scale of the event processes, as mortality and, presumably, UTI recurrences naturally depend on age. Because the participants were between 64 and 102 years old when they entered the study, the observations are then left-truncated.
In our analysis, we seek to assess the effects of the cranberry treatment, to estimate the age-specific rates of UTIs and death, and to determine whether there is an association between UTIs and death risks.

While different joint models for recurrent events and death have been proposed, we focus here on the joint frailty model introduced by \citet{Liuetal:2004}. This model has been applied repeatedly in medical studies (e.g., to study recurrent cancer events as in \citealp{Rondeauetal:2007}, or recurrent heart failure hospitalizations as in \citealp{Rogersetal:2016}), and extended in several directions (e.g., to the setting of nested case-control studies by \citealp{Jazicetal:2019}). The joint frailty model allows us to examine the shape of the rates of recurrence and death, as well as the potential dependence between the two processes. The dependence is introduced by a shared individual random effect entering both the rate of recurrence and the hazard of death. An additional parameter determines whether the processes are positively or negatively associated, and how strong this association is.
In other models, the frailty affects both event rates in the same way \citep{HuangWang:2004}, or the dependence between the processes is left completely unspecified \citep{CookLawless:1997,GhoshLin:2003}.
A common approach to estimating the joint frailty model that we consider here is using Gaussian quadrature to approximate the marginal likelihood \citep{LiuHuang:2008,Rondeauetal:2007}. We will see that in this framework, the setting with left-truncated data can be handled in a straightforward manner by adapting the likelihood.

Incorporating left truncation, which is also called delayed entry, has received varying levels of attention in studies based on different frailty models and joint models.
Several studies have discussed handling left truncation in shared frailty models for clustered survival data \citep[e.g.,][]{Jensenetal:2004,vandenBerg:2016}.
In a recurrent event setting, \citet{Balanetal:2016} considered event dependent selection; i.e., individuals were included in the study only if they had experienced at least one recurrent event in a given time period. Recurrent event studies with selection dependent on survival were briefly discussed in \citet{CookLawless:2007}, but not specifically in the context of frailty models.
It has been argued that for the joint frailty model for recurrent events and death, as specified by \citet{Liuetal:2004}, delayed entry can be easily incorporated \citep{Rondeauetal:2007}. However, to our knowledge, no detailed account of this approach has previously been provided. At the time of writing, the \texttt{R} package \texttt{frailtypack}, which can be used for fitting a variety of frailty models, does not provide functionality for estimating the joint frailty model from left-truncated data (according to the manual of version 3.3.2, date 2020-10-07, \citealp{frailtypack}).
Extensions to left-truncated data have been considered in some other joint models for recurrent events and death. \citet{Emuraetal:2017} introduced a joint frailty-copula model for two event times that can be adapted to accommodate left truncation and recurrent event data. Outside of the class of shared frailty models, \citet{Caietal:2017} proposed a model for longitudinal measurements, recurrent events, and a terminal event with inferences based on estimation equations that can be generalized to allow for left-truncated data. \citet{Liuetal:2012} presented an estimating equation procedure for a partial marginal model of recurrent events in the presence of a terminal event with left truncation.
In the context of joint models for longitudinal data and death, estimation procedures based on left-truncated data have been derived for different models with shared random effects \citep[see, for example,][]{Ardo:2016,Crowtheretal:2016,Piulachsetal:2021}.

In this paper, we propose a method for estimating the joint frailty model for recurrent events and a terminal event, as introduced by \citet{Liuetal:2004}, when data are left-truncated. Our approach adapts the method of \citet{LiuHuang:2008}, who use Gaussian quadrature to approximate the marginal likelihood of the joint frailty model.

The paper is structured as follows.
Section~\ref{sec:model} first presents the joint frailty model and the corresponding likelihood in the usual setting without truncation, and then shows the adjustments for left-truncated data.
The method of estimation is detailed in Section~\ref{sec:estim}, and its performance is assessed via simulation studies in Section~\ref{sec:simu}.
We illustrate the approach using the data set on recurrent UTIs in Section~\ref{sec:appl}, and conclude with a discussion in Section~\ref{sec:disc}.

\section{Joint frailty model and left truncation}\label{sec:model}

The joint frailty model for recurrent events and a terminal event has been applied most frequently in situations in which the time of the terminal event is subject to independent right-censoring only. In the following, we will first present the model and the corresponding likelihood for such right-censored data. Then, we will lay out how certain assumptions and, in particular, the likelihood function are adjusted to the case of left-truncated data. Throughout, we will often refer to the terminal event as death for the sake of simplicity.

\subsection{Joint frailty model}\label{subsec:JFMRC}

We consider independent individuals~$i$, $i=1,...,m$, who can experience recurrent events between time~$t_0=0$ and the time~$D_i$ of the terminal event. Let~$C_i$ denote a censoring time, which is assumed to be independent of the recurrence and terminal event processes. An individual can then be observed only up to his/her follow-up time~$X_i=\min(C_i,D_i)$, and~$\delta_i=\mathbbm{1}\{D_i\leq C_i\}$ will indicate whether the terminal event occurred before censoring, with the indicator function~$\mathbbm{1}\{\cdot\}$. The at-risk indicator at time~$t\geq 0$ is given by $Y_i(t)=\mathbbm{1}\{t\leq X_i\}$, if individuals enter the study at~$t_0=0$.

The number of recurrent events experienced by individual~$i$ up to time~$t$ is recorded by the actual recurrent event process~$N_i^{R\ast}(t)$. Similarly, we define the actual counting process of the terminal event as $N_i^{D\ast}(t)=\mathbbm{1}\{D_i\leq t\}$.
However, due to right-censoring, we can only observe the processes $N_i^D(t)=\mathbbm{1}\{X_i\leq t,\delta_i=1\}$ and $N_i^R(t)=\int_0^t Y_i(s)\mathrm{d}N_i^{R\ast}(s)=N_i^{R\ast}(\min{(t,X_i)})$.
Here, the increment of the recurrence process~$\hbox{d}N_i^{R\ast}(t)=N_i^{R\ast}((t+\hbox{d}t)^-)-N_i^{R\ast}(t^-)$ equals the number of events in the small interval~$[t,t+\hbox{d}t)$, with~$t^-$ as the left-hand limit.

The observed data of individual~$i$ up to time~$t$ are given by $\mathcal{O}_i(t)= \{Y_i(s),N_i^R(s),\\ N_i^D(s), 0\leq s\leq t;\boldsymbol{z}_i\}$, including the observed time-fixed covariate vector~$\boldsymbol{z}_i$. Individual risks will depend on the covariates as well as on the unobservable frailty value~$u_i$, where the frailties~$u_i$ are independent realizations of a positive random variable~$U$.

In the joint frailty model introduced by \citet{Liuetal:2004}, the observed recurrence process is assumed to have the intensity~$Y_i(t)\lambda_i(t|u_i)$ with
\begin{equation}\label{eq:RPRC}
\begin{aligned}
\mathrm{P}(\mathrm{d}N_i^R(t)=1\mid \mathcal{F}_{t^-},D_i\geq t)&=Y_i(t)\lambda_i(t|u_i) \mathrm{d}t\\
\lambda_i(t|u_i)\mathrm{d}t=\mathrm{d}\Lambda_i(t|u_i)&=\mathrm{P}(\mathrm{d}N_i^{R\ast}(t)=1\mid\boldsymbol{z}_i,u_i,D_i\geq t).
\end{aligned}
\end{equation}
Here, $\mathcal{F}_t=\sigma\{\mathcal{O}_i(s),0\leq s\leq t,u_i;i=1,...,m\}$ denotes the $\sigma$-algebra generated by the frailty and the observed data.
The terminal event process is characterized by the intensity~$Y_i(t)h_i(t|u_i)$ with 
\begin{equation}\label{eq:DPRC}
\begin{aligned}
\mathrm{P}(\mathrm{d}N_i^D(t)=1\mid \mathcal{F}_{t^-})&=Y_i(t)h_i(t|u_i) \mathrm{d}t\\
h_i(t|u_i)\mathrm{d}t=\mathrm{d}H_i(t|u_i)&=P(\mathrm{d}N_i^{D\ast}(t)=1\mid \boldsymbol{z}_i, u_i, D_i\geq t).
\end{aligned}
\end{equation}
The first lines in~\eqref{eq:RPRC} and~\eqref{eq:DPRC} follow from the assumption that the censoring mechanism is conditionally independent of the two event processes given the process history.

Following \citet{Liuetal:2004}, we specify the intensities as
\begin{equation}\label{eq:JFM}
\begin{aligned}
\lambda_i(t|u_i)&=u_i\,e^{\boldsymbol{\beta}'\boldsymbol{z}_i} \,\lambda_0(t),\\
h_i(t|u_i)&=u_i^{\gamma}\,e^{\boldsymbol{\alpha}'\boldsymbol{z}_i}\,h_0(t).
\end{aligned}
\end{equation}
The baseline rates of recurrence and death, $\lambda_0(t)$ and $h_0(t)$, are affected by the covariates~$\boldsymbol{z}_i$ through a multiplicative model with effects~$\boldsymbol{\beta}$ and $\boldsymbol{\alpha}$, respectively.
The inclusion of the frailty~$u$ in the recurrence rate accommodates heterogeneity across individuals and dependence between the recurrences within one individual. The association between the recurrent events and death results from the fact that the shared frailty~$u$ also enters the hazard of death. Due to the additional parameter~$\gamma$, the model can capture associations of variable magnitudes and in different directions. For positive $\gamma>0$, individuals with a higher rate of recurrence will also be subject to a higher hazard of death. For $\gamma<0$, a higher rate of recurrence implies a lower hazard of death. If $\gamma=0$, the intensities in~\eqref{eq:JFM} do not share any parameters, and the censoring of the recurrence process by death is non-informative.

A common choice for the distribution of the frailty~$U$ is a gamma distribution with a mean of one and a variance of~$\theta$. We will more generally consider that the frailties~$u_i$ follow a distribution with a density of~$g_{\theta}(u)$ and a corresponding distribution function~$G_{\theta}(u)$ with parameter~$\theta$.
This assumption refers to the initial distribution of frailties in the population at time~$t_0=0$. 
However, if $\gamma\neq 0$, the distribution of frailties in the population will change over time due to selection effects,  which will cause the population at time~$t$ to be composed of survivors with lower mortality risks.

We now formulate the likelihood of the joint frailty model~\eqref{eq:JFM} when individuals are observed from time~$t_0=0$. Let $t_{ij}$, $j=1,...,J_i$, be the observed recurrence times of individual~$i$. Based on the arguments stated in \citet{Liuetal:2004}, the conditional likelihood contribution of individual~$i$ given his/her frailty value~$u_i$ can be written as
\begin{align}\label{eq:LLcompl}
L_i^{(c)}(u_i)&=\left[\prod_{j=1}^{J_i}\lambda_i(t_{ij}|u_i)\right]\,\exp{\left\{-\int_0^{\infty}Y_i(s)\lambda_i(s|u_i)\,\mathrm{d}s\right\}}\notag\\
&\hspace{2cm}\cdot h_i(x_i|u_i)^{\delta_i}\,\exp{\left\{-\int_0^{\infty}Y_i(s)h_i(s|u_i)\,\mathrm{d}s\right\}}\notag\\
&=\left[\prod_{j=1}^{J_i}u_i e^{\boldsymbol{\beta}'\boldsymbol{z}_i}\lambda_0(t_{ij})\right]\,\exp{\left\{-\int_{0}^{x_i}u_i e^{\boldsymbol{\beta}'\boldsymbol{z}_i}\lambda_0(s)\,\mathrm{d}s\right\}}\\
&\hspace{2cm}\cdot \left[u_i^{\gamma} e^{\boldsymbol{\alpha}'\boldsymbol{z}_i} h_0(x_i)\right]^{\delta_i}\,\exp{\left\{-\int_{0}^{x_i} u_i^{\gamma} e^{\boldsymbol{\alpha}'\boldsymbol{z}_i} h_0(s)\,\mathrm{d}s\right\}}.\notag
\end{align}
The marginal likelihood~$L_i$ of the observed data of individual~$i$ is obtained by integrating the above expression over the frailty distribution,
\begin{equation}\label{eq:intF}
L_i=\int_0^{\infty} L_i^{(c)}(u)\,\mathrm{d}G_{\theta}(u)=\int_0^{\infty} L_i^{(c)}(u)\,g_{\theta}(u)\,\mathrm{d}u.
\end{equation}

\subsection{Adjusting for left truncation}\label{subsec:JFMwLT}

We will now extend the above framework to allow for left truncation, that is, individuals entering the study at times that may be later than $t_0=0$. Before deriving the likelihood for the left-truncated data, we introduce some additional notations and assumptions.

A sample of $m_V$ independent individuals~$i$, $i=1,...,m_V$, is left-truncated if the individuals~$i$ enter the study only at times~$V_i\geq t_0$, with strict inequality for at least some individuals. Then, the observation of individual~$i$ is conditional on his/her survival up to the entry time, $D_i>V_i$, and events can only be observed in the interval $[V_i,X_i]$. Hence, the at-risk indicator~$Y_i(t)$ of Section~\ref{subsec:JFMRC} is replaced by ${}_V\!Y_i(t)=\mathbbm{1}\{V_i\leq t \leq X_i\}$.

As a consequence, the observed recurrent event process~${}_V\!N_i^R(t)=\int_0^t {}_V\!Y_i(s)\mathrm{d}N_i^{R\ast}(s)\\ =[N_i^{R\ast}(\min{(t,X_i)})-N_i^{R\ast}(V_i)]\mathbbm{1}\{t>V_i\}$ in this setting records only the recurrences after study entry at $V_i$. Analogously, the left-truncated counting process of the terminal event is given by ${}_V\!N_i^D(t)=\mathbbm{1}\{V_i\leq X_i\leq t, \delta_i=1\}$.
The observed data for individual~$i$ are then ${}_V\!\mathcal{O}_i(t)=\{{}_V\!Y_i(s),{}_V\!N_i^R(s),{}_V\!N_i^D(s), V_i\leq s\leq t;\boldsymbol{z}_i;V_i\}$, and the $\sigma$-algebra is modified as ${}_V\!\mathcal{F}_t=\sigma\{{}_V\!\mathcal{O}_i(s), 0\leq s\leq t, u_i; i=1,...,m_V\}$.

In addition to the assumption of conditionally independent censoring already made in Section~\ref{subsec:JFMRC}, we further assume that the truncation times~$V_i$ are conditionally independent of the recurrence and terminal event processes given the process history. Hence, the intensity of the observed recurrence process is given by ${}_V\!Y_i(t)\lambda_i(t|u_i)$ and \eqref{eq:RPRC} is adapted as
\begin{equation}\label{eq:RPLT}
\mathrm{P}(\mathrm{d}\,{}_V\!N_i^R(t)=1\mid {}_V\!\mathcal{F}_{t-},D_i\geq t)={}_V\!Y_i(t)\lambda_i(t|u_i).\tag{\ref{eq:RPRC}'}
\end{equation}
The intensity of the observed terminal event process is, correspondingly, ${}_V\!Y_i(t) h_i(t|u_i)$, such that~\eqref{eq:DPRC} is modified as
\begin{equation}\label{eq:DPLT}
\mathrm{P}(\mathrm{d}\,{}_V\!N_i^D(t)=1\mid {}_V\!\mathcal{F}_{t-})={}_V\!Y_i(t) h_i(t|u_i).\tag{\ref{eq:DPRC}'}
\end{equation}

Based on this, we can develop the likelihood of the joint frailty model~\eqref{eq:JFM} for left-truncated data.
The conditional likelihood contribution of individual~$i$ given $u_i$ is constructed in analogy to~\eqref{eq:LLcompl}, with $Y_i(s)$ replaced by ${}_V\!Y_i(s)$, and appropriately restricting to the observed recurrence times~$t_{ij}\geq v_i$; that is,
\begin{equation}\label{eq:condLLT}
\begin{aligned}
{}_V\!L_i^{(c)}(u_i)&=\left[\prod_{t_{ij}\geq v_i}\lambda_i(t_{ij}|u_i)\right]\,\exp{\left\{-\int_0^{\infty} {}_V\!Y_i(s)\lambda_i(s|u_i)\,\mathrm{d}s\right\}}\\
&\hspace{2cm}\cdot h_i(x_i|u_i)^{\delta_i}\,\exp{\left\{-\int_0^{\infty} {}_V\!Y_i(s)h_i(s|u_i)\,\mathrm{d}s\right\}}\\
&=\left[\prod_{t_{ij}\geq v_i} u_i e^{\boldsymbol{\beta}'\boldsymbol{z}_i}\lambda_0(t_{ij})\right]\,\exp{\left\{-\int_{v_i}^{x_i}u_i e^{\boldsymbol{\beta}'\boldsymbol{z}_i}\lambda_0(s)\,\mathrm{d}s\right\}}\\
&\hspace{2cm}\cdot \left[u_i^{\gamma} e^{\boldsymbol{\alpha}'\boldsymbol{z}_i} h_0(x_i)\right]^{\delta_i}\,\exp{\left\{-\int_{v_i}^{x_i} u_i^{\gamma} e^{\boldsymbol{\alpha}'\boldsymbol{z}_i} h_0(s)\,\mathrm{d}s\right\}}.
\end{aligned}
\end{equation}
The marginal likelihood contribution is again obtained by integrating out the frailty. However, as the frailty distribution in the sample of survivors differs from the frailty distribution at time $t_0$, we need to integrate over the conditional frailty distribution given survival to the time of entry into the study. This point has previously been discussed in the context of clustered survival data, among others, by \citet{vandenBerg:2016} and \citet{Erikssonetal:2015} and for general state duration models by \citet{LawlessFong:1999}. More formally, the marginal likelihood contribution of individual~$i$ is thus
\begin{equation}\label{eq:margLLT1}
{}_V\!L_i=\int_0^{\infty} {}_V\!L_i^{(c)}(u)\,\mathrm{d}G_{\theta}(u\mid D_i> v_i,V_i=v_i,\boldsymbol{z}_i)=\int_0^{\infty} {}_V\!L_i^{(c)}(u)\,\mathrm{d}G_{\theta}(u\mid D_i> v_i,\boldsymbol{z}_i),
\end{equation}
under the assumption that the truncation time~$V_i$ is independent of~$u$.

In particular, if $\gamma>0$ such that the recurrence process and the mortality process are positively associated, individuals who survived up to time $v$ will tend to have lower frailty values than individuals who died before time~$v$, for given~$\boldsymbol{z}_i$. Hence, the frailty distribution among survivors beyond time $v$, $G_{\theta}(u\mid D>v)$, will tend to have more probability mass at lower values~$u$ than the frailty distribution~$G_{\theta}(u)$ in the underlying population at time~$t_0$.
Consequently, neglecting the effect of the survivor selection on the frailty distribution in the sample and constructing a marginal likelihood as
\begin{equation}\label{eq:margLnaive}
{}_V\!L_i^{\text{naive}}=\int_0^{\infty} {}_V\!L_i^{(c)}(u)\,\mathrm{d}G_{\theta}(u)=\int_0^{\infty} {}_V\!L_i^{(c)}(u)\,g_{\theta}(u)\,\mathrm{d}u,
\end{equation}
would lead to an invalid inference if $\gamma\neq 0$. We will illustrate the resulting biases in the parameter estimates in a simulation study in Section~\ref{sec:simu}.

For computing the correct marginal likelihood~\eqref{eq:margLLT1}, we first apply Bayes' theorem to find that
\begin{equation}\label{eq:Bayes}
g_{\theta}(u| D_i>v_i,\boldsymbol{z}_i)=\frac{\mathrm{P}(D_i> v_i\mid u,\boldsymbol{z}_i)\,g_{\theta}(u)}{\mathrm{P}(D_i> v_i\mid \boldsymbol{z}_i)}=\frac{\exp{\left\{-\int_0^{v_i} h_i(s|u)\,\mathrm{d}s\right\}}\,g_{\theta}(u)}{\int_0^{\infty} \mathrm{P}(D_i> v_i\mid u)\,g_{\theta}(u)\,\mathrm{d}u},
\end{equation}
where we suppress the dependence on the covariates~$\boldsymbol{z}_i$ in the last expression for notational convenience.
Combining equations~\eqref{eq:condLLT}, \eqref{eq:margLLT1}, and \eqref{eq:Bayes}, we can express the marginal likelihood contribution~${}_V\!L_i$ of individual~$i$ with left-truncated data as
\begin{equation}\label{eq:margLLT}
\frac{\int_0^{\infty} \left[\prod_{t_{ij}\geq v_i}\lambda_i(t_{ij}|u)\right]\,\exp{\left\{-\int_{v_i}^{x_i} \lambda_i(s|u)\,\mathrm{d}s\right\}}\,h_i(x_i|u)^{\delta_i}\exp{\left\{-\int_0^{x_i} h_i(s|u)\,\mathrm{d}s\right\}}\,g_{\theta}(u)\,\mathrm{d}u}{\int_0^{\infty} \mathrm{P}(D_i> v_i\mid u)\,g_{\theta}(u)\,\mathrm{d}u}.
\end{equation}

Interestingly, the formula for~${}_V\!L_i$ in \eqref{eq:margLLT} could have been equivalently derived as the marginal (with respect to the frailty) probability of the recurrence and follow-up data on individual~$i$, conditional on individual~$i$ being included in the study, $D_i> v_i$. To see this, let us denote by~$E_i$ the event that individual~$i$ has follow-up time~$x_i$ with indicator~$\delta_i$ and the observed recurrence times~$t_{ij}$ over $[v_i,x_i]$, and consider
\[\mathrm{P}(E_i\mid D_i > v_i)=\frac{\mathrm{P}(E_i\cap \{D_i > v_i\})}{\mathrm{P}(D_i > v_i)}=\frac{\mathrm{P}(E_i)}{\mathrm{P}(D_i > v_i)}=\frac{\int_0^{\infty} \mathrm{P}(E_i\mid u)\,g_{\theta}(u)\,\mathrm{d}u}{\int_0^{\infty}\mathrm{P}(D_i > v_i\mid u)\,g_{\theta}(u)\,\mathrm{d}u}.\]

As the integrals over the frailty distribution in~\eqref{eq:margLLT} will not, in general, have closed form expressions, we will use numerical integration in the following.

\section{Estimation of the joint frailty model}\label{sec:estim}

\citet{LiuHuang:2008} proposed using Gaussian quadrature to approximate the marginal likelihood of frailty proportional hazards models, including the joint frailty model~\eqref{eq:JFM}. In combination with a piecewise constant specification of the baseline rates, this approach allows for the direct maximization of the approximated likelihood. 

The aim of Gaussian quadrature is to replace the integral of a function with a weighted sum of function values. The Gauss-Hermite quadrature rule gives an approximation for a specific integral of a function~$f(x)$,
\[\int_{-\infty}^{\infty} f(x) e^{-x^2}\mathrm{d}x\approx \sum_{q=1}^Q w_q f(x_q).\]
The quadrature points~$x_q$ need to be determined as the roots of the $Q^{\mathrm{th}}$-order Hermite polynomial, while $w_q$ specify corresponding weights.
As the marginal likelihood of a model with normal random effects is easily rewritten in the above form, it follows that such a likelihood can be approximated as
\begin{equation}\label{eq:GHnorm}
\int_{-\infty}^{\infty} L^{(c)}(b)\,\phi(b)\,\mathrm{d}b \approx \sum_{q=1}^Q \tilde{w}_q\, L^{(c)}(\tilde{x}_q)\,\phi(\tilde{x}_q),
\end{equation}
where $\phi(\cdot)$ denotes the standard normal density, while $\tilde{x}_q=\sqrt{2}x_q$ and $\tilde{w}_q=\sqrt{2}w_q e^{x_q^2}$ are modified quadrature points and weights, respectively.
Marginal likelihoods integrated over non-normal random effects can be expressed as an integral of the form in~\eqref{eq:GHnorm} by applying the probability integral transformation \citep[see][]{Nelsonetal:2006,LiuHuang:2008}.
We show in Section~S.1 of the supplementary material how this quadrature approach can be used to derive an approximation of the marginal likelihood of the joint frailty model with left truncation.

The evaluation of the approximate marginal likelihood depends on the specific form of the baseline rates~$\lambda_0(t)$ and $h_0(t)$. As suggested by \citet{LiuHuang:2008}, we adopt a piecewise constant model for these functions,
\[\lambda_0(t)=\sum_{k=1}^{K^R} \lambda_{0k}\mathbbm{1}\{t\in I_k^R\}\quad\text{and}\quad h_0(t)=\sum_{k=1}^{K^D} h_{0k}\mathbbm{1}\{t\in I_k^D\}.\]
with intervals $I_k^R=(t_{k-1}^R,t_k^R]$, $k=1,...,K^R$, and $I_k^D=(t_{k-1}^D,t_k^D]$, $k=1,...,K^D$. Specifications with a moderate number of up to 10 intervals and the cut-points~$t_k$, $k\geq 1$, which have been determined based on the quantiles of the observed event times, are generally expected to lead to good results in practice \citep[see][]{CookLawless:2007,LiuHuang:2008}. In the setting with left truncation, appropriate choices have to be made for the starting points of the first intervals, $t_0^R$ and $t_0^D$. Depending on the study design, they might be set equal to the lowest study entry time, $\min_i v_i$, or a lower value~$t^{\ast}\geq t_0$.

The direct maximization of the marginal likelihood would also be possible if the baseline rates were assumed to follow a simple parametric model, such as the Weibull model. Nonetheless, we recommend the use of the more flexible piecewise constant rate models, unless prior knowledge allows for an informed choice of a specific parametric model.

Finally, the parameter estimates in the joint frailty model with left-truncated data are obtained by maximizing the approximate marginal log-likelihood. The calculation of the standard errors is based on the inverse of the negative Hessian matrix of the approximate marginal log-likelihood. We give additional details on the implementation in Section~S.2 of the supplementary material.

\section{Simulation study}\label{sec:simu}

To evaluate the performance of the proposed method for estimating the parameters of the joint frailty model in case of left-truncated data, we conducted a simulation study. We will also demonstrate which biases can arise if the likelihood is not correctly adjusted to the survivor selection, in particular, to the selection effects on the frailty distribution.

Estimator performance will depend on various aspects of the observation scheme. One aspect is the distribution of the study entry times~$V_i$, in which both the range of the distribution and its shape matter. Furthermore, the censoring mechanism -- that is, the length of the individual follow-up periods and the number of additional drop-outs -- will influence the performance of the method.
To study these issues, we will first present a base scenario, and will then assess how different observational settings affect the results.

\subsection{Settings}

In the base scenario, we generated data from a joint frailty model~\eqref{eq:JFM}. The time scale $t$ is the age of the individual. The hazard of death and the rate of recurrence are each affected by a single binary covariate, which is drawn from a Bernoulli distribution with parameter $0.5$. The regression coefficients are~$\alpha=0.5$ (death) and $\beta=0.5$ (recurrence), respectively. The frailty values are realizations of a gamma distribution with a mean of one and a variance of $\theta=0.5$. The values of the dependence parameter~$\gamma$ were chosen to cover a positive ($\gamma=0.5$) and a negative ($\gamma=-0.5$) association between the recurrence process and death, as well as the case in which the recurrence rate does not affect the mortality risk ($\gamma=0$).

The baseline rates $h_0(t)$ and $\lambda_0(t)$ were designed to mimic a study in an older population among whom the death rates as well as the recurrent event rates increase exponentially with age. Hence, we chose for both baseline rates the Gompertz-Makeham form, $a e^{b t}+c$, where $t=0$ corresponds to age~75. By setting $a=0.984$, $b=0.045$, and $c=0$ for the recurrence process ($\lambda_0(t)$), and $a=0.108$, $b=0.07$, and $c=0.12$ for the survival process ($h_0(t)$), the baseline rates were comparable to the estimated rates for the high risk group in the data example in Section~\ref{sec:appl}.

To arrive at the left-truncated samples, the following steps were combined. For each individual, a survival time $D$ (i.e.,~age $>75$) and an entry time into study $V$ were simulated. Only those individuals who survived beyond his/her entry time -- that is, for whom $D >V $ -- were included in the final sample (i.e., were 'observed'). Therefore, the distribution of entry times~$V$ that are observed in the final sample depends on both the mortality model in~\eqref{eq:JFM} and the initial distribution of the truncation times before selection.

In the base scenario, our aim was to have entry times in the final sample that were distributed across the total age range -- here, ages $75$ to $95$ -- with higher numbers of study entries at the younger ages than at the older ages. This scheme will be referred to as truncation pattern~A in the following.

\begin{figure}[htb]
\centering
\makebox{\includegraphics[width=\textwidth]{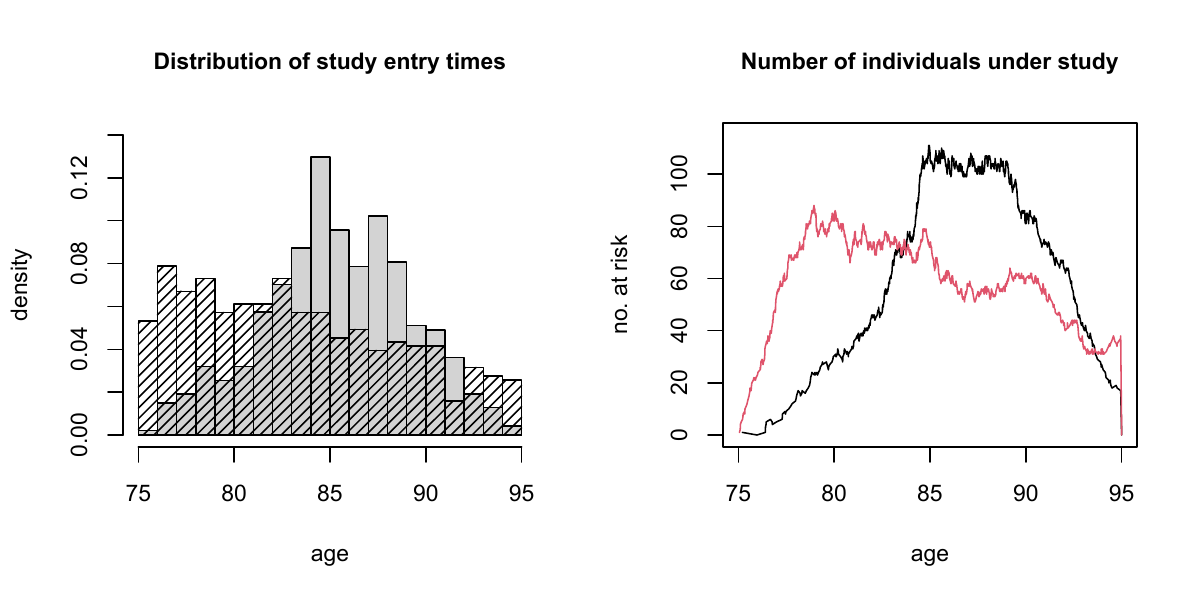}}
\caption{Distribution of the ages at study entry (left) and the number of individuals at risk across the age range $[75,95]$ (right) for one simulated sample from the base scenario with truncation pattern~A (shaded bars, red line) or truncation pattern~B (gray bars, black line), both with planned individual follow-ups of four years.}
\label{fig:simuAtRisk}
\end{figure}

To obtain a final observed sample with such characteristics, the entry times~$V$ were drawn from a truncated normal distribution defined on the age range 75 to 95.
More specifically, the truncated normal distribution was specified to have a mode equal to the maximum age of 95 with parameter values chosen so that the distribution of the observed study entry times in the truncated sample had the desired shape (the left panel of Figure~\ref{fig:simuAtRisk} illustrates this procedure). The initial number of generated survival times was chosen such that the final truncated samples had an average size of about $m_V=500$ individuals.

An independent censoring mechanism was imposed in the following way. For most individuals, the censoring times were the end of a planned individual follow-up period of $t_C=4$ years. However, some of the follow-up times were longer than four years, and some premature random drop-outs occurred. Again, this was done in response to the situation that we observed in the data application of Section~\ref{sec:appl}.
Accordingly, we generated random durations from a mixture distribution with an 85\% point mass at $t_C$, a 10\% uniform distribution on $[0,t_C]$, and a 5\% uniform distribution on $[t_C,t_C+0.5]$, with the latter two covering the drop-outs before $t_C$ and the longer follow-up periods, respectively. These random durations were added to the individual~$V_i$, and the individual censoring time~$C_i$ was the minimum of this sum and age~95.

The right panel of Figure~\ref{fig:simuAtRisk} illustrates how the mechanisms of truncation and censoring jointly determine the number of individuals at risk at any time~$t$ across the age range $[75,95]$. Truncation pattern~A causes the number of individuals at risk to increase steeply at the early ages, and then to decrease only gradually across the age range. However, due to the relatively short individual follow-up times, the number of individuals at risk across ages is considerably smaller than the total sample size of about 500.

In the setting with a positive association between the recurrence and mortality process ($\gamma = 0.5$), additional simulation scenarios were set up by varying the censoring and truncation patterns. 

First, we considered the effect of changing the planned individual follow-up times to $t_C=1$ year or $t_C=8$ years, respectively. Longer individual follow-up times increase the number of individuals who are under observation at a certain time~$t$, and are therefore expected to improve the estimator performance.

Second, we explored a scenario with a more unimodal distribution of the study entry times in which relatively few individuals entered the study at the youngest and the oldest ages (see Figure~\ref{fig:simuAtRisk}). This is truncation pattern~B. To obtain a final sample with these characteristics, we simulated the initial truncation times again from a truncated normal distribution on the age range 75 to 95. However, in this scenario, the distribution had a mode equal to 90, that is, within the above age range.

Finally, we examined a setting with a wider age range of~$[64,105]$. If $t=0$ was now expected to correspond to age 64, but the Gompertz-Makeham rates were expected to agree with the rates of the base scenario over  $[75,95]$, the parameters needed to be adapted. This was achieved by maintaining the values of $b$ and $c$, but setting $a=0.6$ or $a=0.05$ for the recurrence and death processes, respectively. The initial distributions of the study entry times were adapted to produce truncation patterns A or B on the wider age range $[64,105]$. 
In all of the additional scenarios, the truncated samples again had a target size of $m_V=500$ individuals.
The parameter values for the distributions of the study entry times and the initial sample sizes for the different scenarios can be found in Section~S.3 of the supplementary material.

\subsection{Estimation and results}

The estimation of the joint frailty model was carried out under the assumptions of gamma-distributed frailties with a mean of one and piecewise constant models for the two baseline rates $\lambda_0(t)$ and $h_0(t)$. For both rates, 10 intervals were used that were denoted by~$I_k^R$ (recurrence process) and $I_k^D$ (mortality), $k=1,...,10$. The intervals were determined by the deciles of the observed recurrence and survival times, respectively.
We set $t_0^R=t_0^D=75$ (or $t_0^R=t_0^D=64$) equal to the starting point of the respective age range and $t_{10}^R=t_{10}^D$ equal to the maximum follow-up time in the sample. The marginal likelihood was approximated using non-adaptive Gauss-Hermite quadrature with $Q=30$ quadrature points. We ran 200 replications in each setting. All computations were performed in R \citep{R}. Further details on the implementation are provided in Section~S.2 of the supplementary material.

\begin{figure}[htb]
\centering
\makebox{\includegraphics[width=\textwidth]{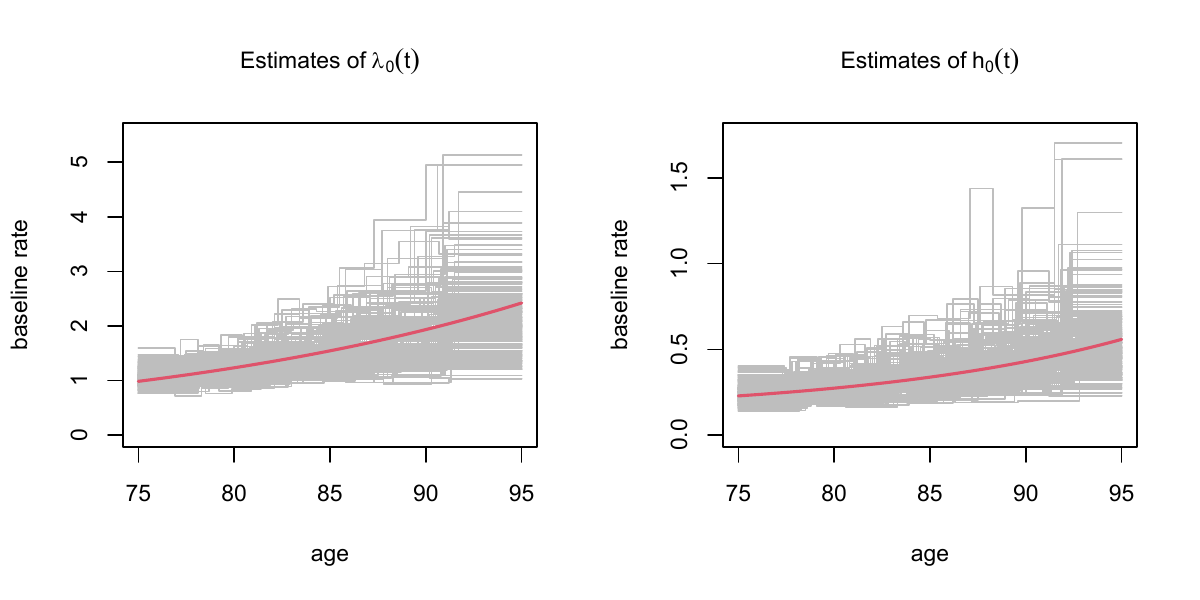}}
\caption{Estimates (gray) of the baseline rate of recurrence (left) and of death (right) based on 200 truncated samples with a target size of $500$ generated from a joint frailty model with positive dependence ($\gamma=0.5$) under the base scenario. The red bold line gives the true baseline rate.}
\label{fig:simuRates}
\end{figure}

Figures~\ref{fig:simuEstim} and \ref{fig:simuRates} illustrate the results of the base scenario with different underlying associations, $\gamma\in\{-0.5,0,0.5\}$. The top panels of Figure~\ref{fig:simuEstim} show that the covariate effects~$\alpha$ and $\beta$, the dependence parameter~$\gamma$, and the frailty variance~$\theta$ are estimated without significant bias. The estimated standard errors of these parameters in the bottom panels of Figure~\ref{fig:simuEstim} are largely in line with the empirical standard deviations of the respective parameter estimates across the replications.
Nevertheless, we notice that the estimator performance varies for different true values of the dependence parameter.

This pattern can be explained to some extent by different survivor selection effects. The truncated sample consists of survivors, who tend to have lower mortality risks. If the recurrence process and the mortality process are positively associated, this implies that the frailty values and the recurrence rates are lower in the sample of survivors. In the current setting, this lower frailty variance in the sample is favorable for the estimation of~$\theta$; whereas the low recurrence rate, which is associated with higher probabilities of having no observed recurrent event, increases the variability in the corresponding estimated covariate effect~$\hat{\beta}$.
The opposite effects are observed if the event processes are negatively associated. If the recurrence rate has no effect on survival ($\gamma=0$), the method still yields reliable results, and the parameters exclusively affecting survival are estimated with higher levels of precision.

The estimates of the baseline rates, displayed for the base scenario with positive dependence~$\gamma=0.5$ in Figure~\ref{fig:simuRates}, also perform satisfactorily. 

\begin{figure}[htb]
\centering
\makebox{\includegraphics[width=\textwidth]{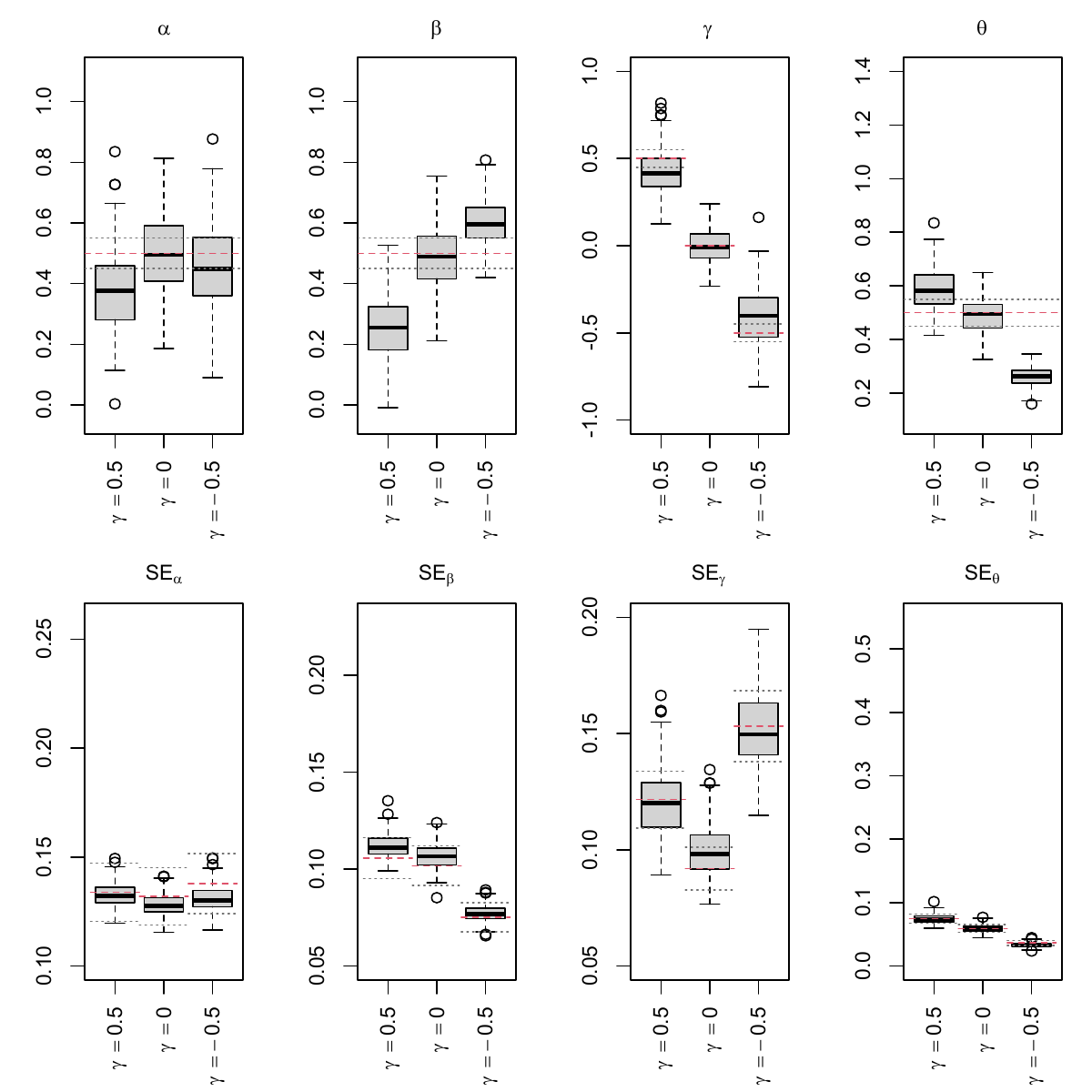}}
\caption{Box plots of the parameter estimates (top) and estimated standard errors (bottom) based on the naive likelihood of the joint frailty model for positive ($\gamma=0.5$), no ($\gamma=0$), or negative ($\gamma=-0.5$) dependence under the base scenario. Left to right: covariate effect on mortality ($\alpha$) and on recurrences ($\beta$), dependence parameter ($\gamma$), and frailty variance ($\theta$) based on 200 truncated samples with a target size of $500$. The red dashed line marks the true parameter value (top) or empirical standard deviation (bottom); the gray dotted lines mark 10\% deviations from the respective value.}
\label{fig:simuEstimInc}
\end{figure}

\begin{figure}[htb]
\centering
\makebox{\includegraphics[width=\textwidth]{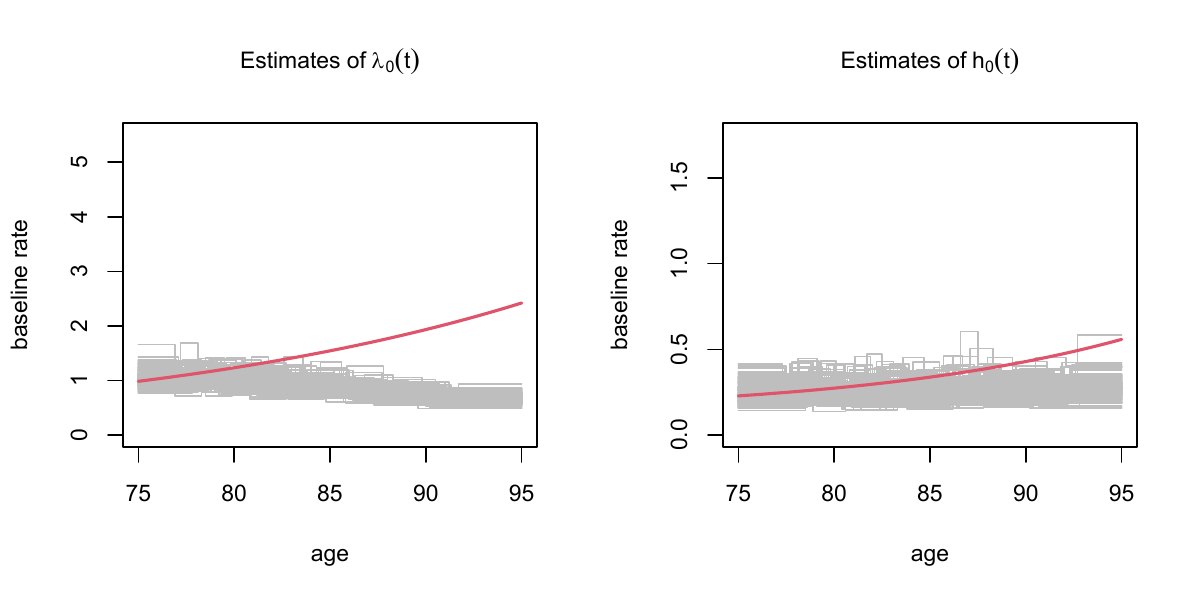}}
\caption{Estimates (gray) of the baseline rate of recurrence (left) and of death (right) based on the naive likelihood for 200 truncated samples with a target size of $500$ generated from a joint frailty model with positive dependence ($\gamma=0.5$) under the base scenario. The red bold line gives the true baseline rate.}
\label{fig:simuRatesInc}
\end{figure}

It is instructive to look at how the results change if the effects of survivor selection on the frailty distribution in the sample are not taken into account correctly.
As Figures~\ref{fig:simuEstimInc} and \ref{fig:simuRatesInc} show, if the inference is based on the naively constructed marginal likelihood~\eqref{eq:margLnaive}, biases can be seen in all parameter estimates in case the recurrence process and the mortality process are associated. Moreover, as the estimated standard errors for the covariate effects are substantially smaller than those obtained using the correct likelihood, they do not adequately reflect the uncertainty in the parameter estimates. The baseline rates of recurrence and death are increasingly underestimated for advancing age in the base scenario with positive dependence ($\gamma=0.5$), as depicted in Figure~\ref{fig:simuRatesInc}. This is because in a setting with a positive association, the distribution of frailty among the survivors tends to be concentrated at lower values. Accordingly, for negative associations, the recurrence rate will be overestimated at the older ages, while the hazard of death will again be underestimated at the older ages. Hence, failing to construct the marginal likelihood based
on the correct distribution of the frailty, see~(\ref{eq:margLLT1}), introduces marked biases in the estimates and the standard errors.
Only if the event rates are not associated ($\gamma=0$) is the distribution of frailty among survivors equal to the initial frailty distribution~$G_{\theta}$, such that the naive marginal likelihood coincides with the correct marginal likelihood~\eqref{eq:margLLT1} and yields valid inferences.

Lastly, we want to examine the results for the additional simulation scenarios with modified censoring and truncation patterns. The figures illustrating these results can be found in Section~S.3 of the supplementary material.
In the scenario with a planned individual follow-up of only $t_C=1$ year, we find increased variability in all parameter estimates (see Figures~S.1 and S.2). This is expected, because with shorter individual follow-up times, fewer individuals are observed at a given age~$t$ than in the base scenario. Further extending the planned individual follow-up times of the base scenario from $t_C=4$ to $t_C=8$ years does not lead to considerable improvements, apart from some reduced variability in the estimates of the frailty variance and the baseline rates.

A change in the distribution of the study entry times can markedly influence the estimation results. In the modified base scenario with truncation pattern~B, the estimated covariate effects~$\hat{\alpha}$ and $\hat{\beta}$ are more variable than under truncation pattern~A, occasionally with negative estimates (see Figure~S.3). In addition, the first piece of each of the baseline rates shows an upward bias (cf.~top panels of Figure~S.4) because few individuals entered the study at the younger ages. Although the intervals for the rate pieces were constructed to contain roughly equal numbers of observed events, the first intervals cover a relative large age range with few individuals under study at a given age~$t$ due to the delayed entry.

The last scenario combines a wider age range~$[64,105]$ and truncation times spread across the whole age range according to pattern~A or~B, with individual follow-ups planned for $t_C=4$ years. This setting is more demanding because the amount of information available at a given age~$t$ is considerably smaller than it is in the scenarios with age range~$[75,95]$. Therefore, the variability in the estimates tends to increase, and the estimates of the dependence parameter and the frailty variance exhibit a small downward bias (see Figure~S.3).

Overall, the simulation studies suggest that the proposed method for the estimation of the joint frailty model based on left-truncated data performs satisfactorily.
The parameter estimates are largely unbiased if the study design ensures that a reasonable number of individuals are under observation across time~$t$. Including a relatively large number of individuals early on and a preferably stable number of study entries across the remaining time range benefits the estimation. In addition, the individual follow-up times should be sufficiently long given the total time window and the sample size. As expected, the patterns of censoring and truncation that cause more information to be lost negatively affect the estimator performance.

\section{Recurrent infections and mortality in an older population}\label{sec:appl}

We use the proposed method to analyze recurrent urinary tract infections and mortality in an institutionalized elderly population. The data come from a double-blind, randomized, placebo-controlled trial in long-term care facilities that aimed to assess the effect of cranberry capsules on the occurrence of UTIs in vulnerable older persons \citep{Caljouw:2014}. At baseline, the participants were stratified into two groups of high or low UTI risk depending on whether they had diabetes mellitus, a urinary catheter, or at least one treated UTI in the preceding year. Within these two strata, the participants were randomly assigned to the treatment or the control group. The participants took cranberry or placebo capsules twice a day over a period of one year.
Occurrences of UTIs were recorded by the treating physicians according to a clinical definition based on international practice guidelines for LTCF residents, and a strict definition based on scientific criteria.
We focus here on the occurrence of the more broadly defined clinical UTIs.

\begin{figure}[htb]
\centering
\makebox{\includegraphics[width=\textwidth]{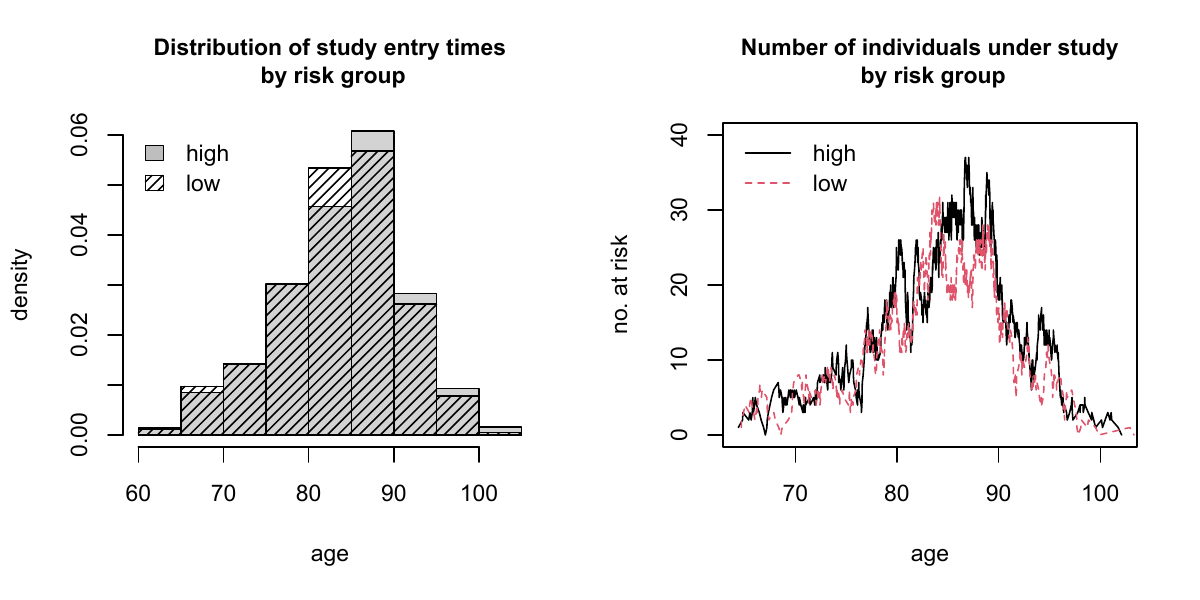}}
\caption{Distribution of the ages at study entry (left) and the number of individuals under observation across the age range (right) in the cranberry data set, separately for the groups with high baseline UTI risk (gray bars, black solid line) and low baseline UTI risk (shaded bars, red dashed line).}
\label{fig:cranbAtRisk}
\end{figure}

The final study population consisted of $928$ individuals, most of whom were women (703; 75.8\%). Of these individuals, $516$ were considered to be at high baseline UTI risk, while $412$ were considered to be at low baseline UTI risk. Individuals entered the study between ages 64 and 102, as shown in the left panel of Figure~\ref{fig:cranbAtRisk}, and were followed on average for about a year (mean: 332 days, median: 372 days). A total of 317 participants ($34.2\%$) died during the study period. The number of observed UTIs per individual ranged from zero to 10, with $62.2\%$ of the individuals having no UTIs, $20.8\%$ having one UTI, and $17.0\%$ experiencing two or more UTIs during the follow-up period.

Unlike in the original study, we modeled recurrent UTIs and mortality to evolve with age, where $t_0=0$ corresponded to age 64. Because of the specific distribution of the ages at study entry in conjunction with the short individual follow-up times, relatively few individuals were under observation at a given age, in particular at the youngest and oldest ages, as the right panel of Figure~\ref{fig:cranbAtRisk} shows.

We estimated the joint frailty model for UTIs and overall mortality with age as the time scale separately for the groups with high and low baseline UTI risk. Two binary covariates for treatment and gender were included, and frailties were assumed to follow a gamma distribution with a mean of one. The baseline rate of UTI recurrence and the hazard of death were specified as piecewise constant functions with 10 intervals over the age range 64 to 103 in the high risk group and 64 to 104 in the low risk group. Separately for the two risk groups, the cut-points for the intervals were determined based on the deciles of the observed recurrence or death times, respectively. The likelihood was approximated using non-adaptive Gaussian quadrature with 30 nodes.

The parameter estimates for both risk groups are reported in Table~\ref{tab:cranb}.
In the group with a high baseline UTI risk, the infection rates varied between participants with an estimated frailty variance of $\hat{\theta}=0.380$ (SE: $0.086$). In particular, individuals with a higher rate of recurrent infections tended to also experience higher mortality risks, as indicated by the positive estimate of the dependence parameter, $\hat{\gamma}=0.181$ (SE: $0.084$).
The participants in the low risk group seemed to be more heterogeneous ($\hat{\theta}=1.122$, SE: $0.316$), but the analysis did not detect an association between the occurrence of UTIs and survival ($\hat{\gamma}=0.058$, SE: $0.044$). The results suggest that the cranberry capsules did not have a noticeable effect on the occurrence of UTIs irrespective of the baseline UTI risk. When we look at gender differences, we see that males and females experienced similar infection rates, while males had higher mortality levels than females in both groups.

\begin{figure}[htb]
\centering
\makebox{\includegraphics[width=\textwidth]{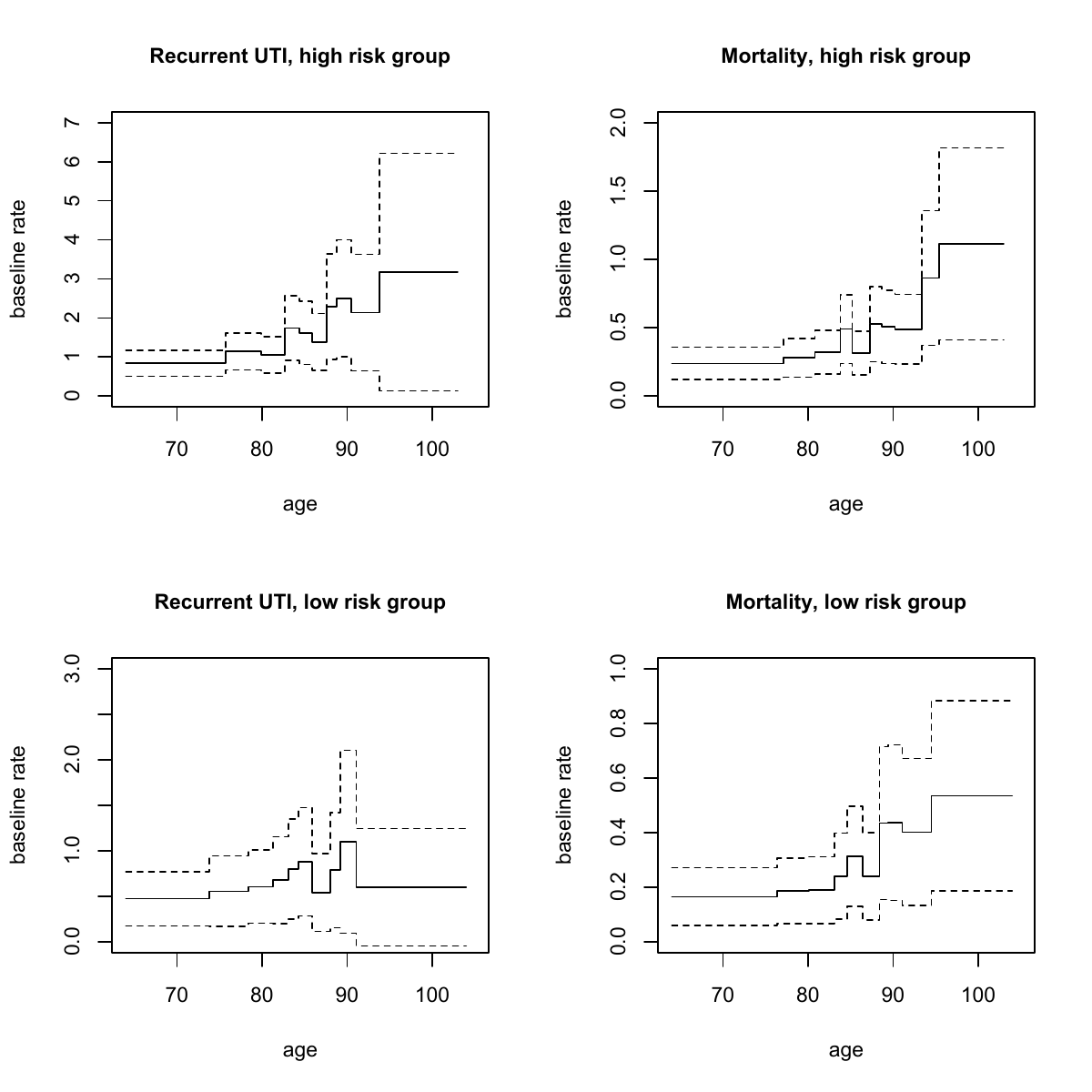}}
\caption{Estimated baseline rates (solid) of recurrence (left) and mortality (right) with $\pm 2$ SE-confidence bounds (dashed) for the cranberry data, separately for the groups with high baseline UTI risk (top) and low baseline UTI risk (bottom).}
\label{fig:cranbRates}
\end{figure}

The estimated baseline rates displayed in Figure~\ref{fig:cranbRates} demonstrate nicely the age dependence of the recurrence rate and the hazard of death.
For the individuals with a high baseline UTI risk, both the rate of recurrent infection and the mortality risk showed a general tendency to increase with age, although the small number of observations leads to considerable uncertainty at the highest ages.
In addition, the individuals with a high baseline UTI risk tended to experience higher rates of recurrent infection and death than the individuals with a low baseline UTI risk.

The original study, which used a different time scale, reported a positive treatment effect of the intake of cranberry capsules only in the group with a high baseline UTI risk and only for the outcome of UTI incidence (first infection during follow-up). When all recurrent UTIs were analyzed in a gamma-frailty model, no treatment effect was detected, which is in line with the findings presented here.

\section{Discussion}\label{sec:disc}

We have proposed a method for estimating the joint frailty model for recurrent events and a terminal event based on left-truncated data.
The marginal likelihood of the model can be expressed as a ratio of two integrals over the frailty distribution, each of which is approximated using Gauss-Hermite quadrature.
The direct maximization of the approximate marginal likelihood is possible if the baseline rates are specified as piecewise constant functions.

The simulation studies presented here have shown that the estimation procedure performs satisfactorily in general, and have demonstrated how different observation schemes affect the estimator performance. While any pattern of truncation or censoring results in incomplete information, study designs should still aim to provide enough information to meet the needs of a model as complex as the joint frailty model. Having a sufficient number of individuals under observation across most of the time range, and especially at the start of the process, seems to be crucial for the method to yield reliable results.

Allowing for left truncation in frailty models requires us to consider carefully how the frailty distribution in the sample of survivors may differ from the frailty distribution in the underlying population due to selection effects. We illustrated through simulations the biases that can arise in the parameter estimates of the joint frailty model if this difference in the frailty distributions is ignored.

Extending the framework of the joint frailty model to incorporate delayed entry allowed us to study age-specific rates of recurrent urinary tract infections and death in an older population.
Similarly, the proposed approach enables researchers to use the joint frailty model in a wider variety of contexts in which subjects are included in a study only if they have not yet experienced the terminal event. Apart from clinical studies with delayed entry, these contexts may include register-based studies of event processes evolving with age as the main time scale, with individuals entering at different ages.

For a complete specification of the model and the approximate likelihood function, we need to choose a frailty distribution as well as the number of quadrature points and the intervals for the baseline rates.
Although the simulation study and the application covered only the common choice of a gamma distribution for the frailties, the quadrature approach can be employed with other frailty distributions that have a closed form inverse distribution function, or a log-normal distribution.
The number of quadrature points then determines the accuracy of the integral approximations in the marginal likelihood, as well as the computation time. In line with previous recommendations for gamma frailty models \citep[see][]{LiuHuang:2008}, we used $Q=30$ quadrature points, which produced good results in a reasonable period of time in our settings.
Regarding the baseline rate functions, the number of intervals corresponds to the number of parameters for the rates, and should thus be selected to allow for sufficient flexibility of the shape of the rates, while retaining numerical stability and the computational feasibility of the method. Adequate results can often be obtained with moderate numbers of up to 10 intervals.

Nevertheless, in some applications, it may seem appealing to aim for smooth rate estimates, such as through the use of penalized splines. However, automatic smoothing parameter selection in joint frailty models is an issue that needs further investigation.

Moreover, the current approach is limited to applications in which there is heterogeneity in recurrence rates. Due to the specific dependence structure in the joint frailty model considered here, the association between the recurrence process and the terminal event process cannot be assessed if the frailty variance is close to zero.

Finally, in some applications, it might be of interest to extend the proposed method to use information on recurrences before entry into the study. These additional event times can be included in the marginal likelihood, and are expected to lead to increased precision in the estimation of the model for the recurrence process. However, when using such an approach, researchers should reflect critically on the quality of the retrospectively collected data, as recollections by study participants may be less reliable than data drawn from other sources, such as registries.

\bibliographystyle{chicago}

\bibliography{JFMwLTbib}

\begin{thebibliography}{}

\bibitem[\protect\citeauthoryear{Balan, Jonker, Johannesma, and Putter}{Balan
  et~al.}{2016}]{Balanetal:2016}
Balan, T.~A., M.~A. Jonker, P.~C. Johannesma, and H.~Putter (2016).
\newblock Ascertainment correction in frailty models for recurrent events data.
\newblock {\em Statistics in Medicine\/}~{\em 35}, 4183--4201.

\bibitem[\protect\citeauthoryear{Cai, Wang, and Chan}{Cai
  et~al.}{2017}]{Caietal:2017}
Cai, Q., M.-C. Wang, and K.~C.~G. Chan (2017).
\newblock Joint modeling of longitudinal, recurrent events and failure time
  data for survivor's population.
\newblock {\em Biometrics\/}~{\em 73}, 1150--1160.

\bibitem[\protect\citeauthoryear{Caljouw, {van den Hout}, Putter, Achterberg,
  Cools, and Gussekloo}{Caljouw et~al.}{2014}]{Caljouw:2014}
Caljouw, M. A.~A., W.~B. {van den Hout}, H.~Putter, W.~P. Achterberg, H.~J.~M.
  Cools, and J.~Gussekloo (2014).
\newblock Effectiveness of cranberry capsules to prevent urinary tract
  infections in vulnerable older persons: A double-blind randomized
  placebo-controlled trial in long-term care facilities.
\newblock {\em Journal of the American Geriatrics Society\/}~{\em 62},
  103--110.

\bibitem[\protect\citeauthoryear{Cook and Lawless}{Cook and
  Lawless}{1997}]{CookLawless:1997}
Cook, R.~J. and J.~F. Lawless (1997).
\newblock Marginal analysis of recurrent events and a terminating event.
\newblock {\em Statistics in Medicine\/}~{\em 16}, 911--924.

\bibitem[\protect\citeauthoryear{Cook and Lawless}{Cook and
  Lawless}{2007}]{CookLawless:2007}
Cook, R.~J. and J.~F. Lawless (2007).
\newblock {\em The Statistical Analysis of Recurrent Events}.
\newblock Statistics for Biology and Health. New York: Springer.

\bibitem[\protect\citeauthoryear{Crowther, Andersson, Lambert, Abrams, and
  Humphreys}{Crowther et~al.}{2016}]{Crowtheretal:2016}
Crowther, M.~J., T.~M.-L. Andersson, P.~C. Lambert, K.~R. Abrams, and
  K.~Humphreys (2016).
\newblock Joint modelling of longitudinal and survival data: incorporating
  delayed entry and an assessment of model misspecification.
\newblock {\em Statistics in Medicine\/}~{\em 35}, 1193--1209.

\bibitem[\protect\citeauthoryear{Emura, Nakatochi, Murotani, and Rondeau}{Emura
  et~al.}{2017}]{Emuraetal:2017}
Emura, T., M.~Nakatochi, K.~Murotani, and V.~Rondeau (2017).
\newblock A joint frailty-copula model between tumour progression and death for
  meta-analysis.
\newblock {\em Statistical Methods in Medical Research\/}~{\em 26}, 2649--2666.

\bibitem[\protect\citeauthoryear{Eriksson, Martinussen, and Scheike}{Eriksson
  et~al.}{2015}]{Erikssonetal:2015}
Eriksson, F., T.~Martinussen, and T.~H. Scheike (2015).
\newblock Clustered survival data with left-truncation.
\newblock {\em Scandinavian Journal of Statistics\/}~{\em 42}, 1149--1166.

\bibitem[\protect\citeauthoryear{Ghosh and Lin}{Ghosh and
  Lin}{2003}]{GhoshLin:2003}
Ghosh, D. and D.~Y. Lin (2003).
\newblock Semiparametric analysis of recurrent events data in the presence of
  dependent censoring.
\newblock {\em Biometrics\/}~{\em 59}, 877--885.

\bibitem[\protect\citeauthoryear{Huang and Wang}{Huang and
  Wang}{2004}]{HuangWang:2004}
Huang, C.-Y. and M.-C. Wang (2004).
\newblock Joint modeling and estimation for recurrent event processes and
  failure time data.
\newblock {\em Journal of the American Statistical Association\/}~{\em 99},
  1153--1165.

\bibitem[\protect\citeauthoryear{Jazi\'c, Haneuse, French, MacGrogan, and
  Rondeau}{Jazi\'c et~al.}{2019}]{Jazicetal:2019}
Jazi\'c, I., S.~Haneuse, B.~French, G.~MacGrogan, and V.~Rondeau (2019).
\newblock Design and analysis of nested case-control studies for recurrent
  events subject to a terminal event.
\newblock {\em Statistics in Medicine\/}~{\em 38\/}(22), 4348--4362.

\bibitem[\protect\citeauthoryear{Jensen, Brookmeyer, Aaby, and Andersen}{Jensen
  et~al.}{2004}]{Jensenetal:2004}
Jensen, H., R.~Brookmeyer, P.~Aaby, and P.~K. Andersen (2004).
\newblock {\em Shared Frailty Model for Left-Truncated Multivariate Survival
  Data}.
\newblock Biostatistisk Afdeling: Museum Tusculanum.

\bibitem[\protect\citeauthoryear{Lawless and Fong}{Lawless and
  Fong}{1999}]{LawlessFong:1999}
Lawless, J.~F. and D.~Y.~T. Fong (1999).
\newblock State duration models in clinical and observational studies.
\newblock {\em Statistics in Medicine\/}~{\em 18}, 2365--2376.

\bibitem[\protect\citeauthoryear{Liu, Schaubel, and Kalbfleisch}{Liu
  et~al.}{2012}]{Liuetal:2012}
Liu, D., D.~E. Schaubel, and J.~D. Kalbfleisch (2012).
\newblock Computationally efficient marginal models for clustered recurrent
  event data.
\newblock {\em Biometrics\/}~{\em 68}, 637--647.

\bibitem[\protect\citeauthoryear{Liu and Huang}{Liu and
  Huang}{2008}]{LiuHuang:2008}
Liu, L. and X.~Huang (2008).
\newblock The use of {G}aussian quadrature for estimation in frailty
  proportional hazards models.
\newblock {\em Statistics in Medicine\/}~{\em 27}, 2665--2683.

\bibitem[\protect\citeauthoryear{Liu, Wolfe, and Huang}{Liu
  et~al.}{2004}]{Liuetal:2004}
Liu, L., R.~A. Wolfe, and X.~Huang (2004).
\newblock Shared frailty models for recurrent events and a terminal event.
\newblock {\em Biometrics\/}~{\em 60}, 747--756.

\bibitem[\protect\citeauthoryear{Modig, Andersson, Drefahl, and Ahlbom}{Modig
  et~al.}{2013}]{Modigetal:2013}
Modig, K., T.~Andersson, S.~Drefahl, and A.~Ahlbom (2013).
\newblock Age-specific trends in morbidity, mortality and case-fatality from
  cardiovascular disease, myocardial infarction and stroke in advanced age:
  Evaluation in the {S}wedish population.
\newblock {\em PLOS ONE\/}~{\em 8}, 1--13.

\bibitem[\protect\citeauthoryear{Nelson, Lipsitz, Fitzmaurice, Ibrahim, Parzen,
  and Strawderman}{Nelson et~al.}{2006}]{Nelsonetal:2006}
Nelson, K.~P., S.~R. Lipsitz, G.~M. Fitzmaurice, J.~Ibrahim, M.~Parzen, and
  R.~Strawderman (2006).
\newblock Use of the probability integral transformation to fit nonlinear
  mixed-effects models with nonnormal random effects.
\newblock {\em Journal of Computational and Graphical Statistics\/}~{\em
  15\/}(1), 39--57.

\bibitem[\protect\citeauthoryear{Piulachs, Andrinopoulou, Guill\'en, and
  Rizopoulos}{Piulachs et~al.}{2021}]{Piulachsetal:2021}
Piulachs, X., E.-R. Andrinopoulou, M.~Guill\'en, and D.~Rizopoulos (2021).
\newblock A {B}ayesian joint model for zero-inflated integers and
  left-truncated event times with a time-varying association: {A}pplications to
  senior health care.
\newblock {\em Statistics in Medicine\/}~{\em 40\/}(1), 147--166.

\bibitem[\protect\citeauthoryear{{R Core Team}}{{R Core Team}}{2020}]{R}
{R Core Team} (2020).
\newblock {\em R: A Language and Environment for Statistical Computing}.
\newblock Vienna, Austria: R Foundation for Statistical Computing.

\bibitem[\protect\citeauthoryear{Rogers, Yaroshinsky, Pocock, Stokar, and
  Pogoda}{Rogers et~al.}{2016}]{Rogersetal:2016}
Rogers, J.~K., A.~Yaroshinsky, S.~J. Pocock, D.~Stokar, and J.~Pogoda (2016).
\newblock Analysis of recurrent events with an associated informative dropout
  time: {A}pplication of the joint frailty model.
\newblock {\em Statistics in Medicine\/}~{\em 35}, 2195--2205.

\bibitem[\protect\citeauthoryear{Rondeau, Gonzalez, Mazroui, Mauguen, Diakite,
  Laurent, Lopez, Kr\'{o}l, Sofeu, Dumerc, and Rustand}{Rondeau
  et~al.}{2020}]{frailtypack}
Rondeau, V., J.~R. Gonzalez, Y.~Mazroui, A.~Mauguen, A.~Diakite, A.~Laurent,
  M.~Lopez, A.~Kr\'{o}l, C.~L. Sofeu, J.~Dumerc, and D.~Rustand (2020).
\newblock {frailtypack}: General frailty models: Shared, joint and nested
  frailty models with prediction; evaluation of failure-time surrogate
  endpoints.
\newblock {R} package version 3.3.2.
\newblock https://CRAN.R-project.org/package=frailtypack.

\bibitem[\protect\citeauthoryear{Rondeau, Mathoulin-Pelissier, Jacqmin-Gadda,
  Brouste, and Soubeyran}{Rondeau et~al.}{2007}]{Rondeauetal:2007}
Rondeau, V., S.~Mathoulin-Pelissier, H.~Jacqmin-Gadda, V.~Brouste, and
  P.~Soubeyran (2007).
\newblock Joint frailty models for recurring events and death using maximum
  penalized likelihood estimation: application on cancer events.
\newblock {\em Biostatistics\/}~{\em 8\/}(4), 708--721.

\bibitem[\protect\citeauthoryear{{van den Berg} and Drepper}{{van den Berg} and
  Drepper}{2016}]{vandenBerg:2016}
{van den Berg}, G.~J. and B.~Drepper (2016).
\newblock Inference for shared-frailty survival models with left-truncated
  data.
\newblock {\em Econometric Reviews\/}~{\em 35}, 1075--1098.

\bibitem[\protect\citeauthoryear{{van den Hout} and Muniz-Terrera}{{van den
  Hout} and Muniz-Terrera}{2016}]{Ardo:2016}
{van den Hout}, A. and G.~Muniz-Terrera (2016).
\newblock Joint models for discrete longitudinal outcomes in aging research.
\newblock {\em Journal of the Royal Statistical Society, Series C\/}~{\em 65},
  167--186.

\end{thebibliography}


\begin{thebibliography}{}

\bibitem[\protect\citeauthoryear{Gilbert and Varadhan}{Gilbert and
  Varadhan}{2019}]{numDeriv}
Gilbert, P. and R.~Varadhan (2019).
\newblock numderiv: Accurate numerical derivatives.
\newblock R package version 2016.8-1.1.
\newblock https://CRAN.R-project.org/package=numDeriv.

\bibitem[\protect\citeauthoryear{Liu and Huang}{Liu and
  Huang}{2008}]{LiuHuang:2008}
Liu, L. and X.~Huang (2008).
\newblock The use of {G}aussian quadrature for estimation in frailty
  proportional hazards models.
\newblock {\em Statistics in Medicine\/}~{\em 27}, 2665--2683.

\bibitem[\protect\citeauthoryear{Mersmann, Trautmann, Steuer, and
  Bornkamp}{Mersmann et~al.}{2018}]{truncnorm}
Mersmann, O., H.~Trautmann, D.~Steuer, and B.~Bornkamp (2018).
\newblock truncnorm: Truncated normal distribution.
\newblock R package version 1.0-8.

\bibitem[\protect\citeauthoryear{Nelson, Lipsitz, Fitzmaurice, Ibrahim, Parzen,
  and Strawderman}{Nelson et~al.}{2006}]{Nelsonetal:2006}
Nelson, K.~P., S.~R. Lipsitz, G.~M. Fitzmaurice, J.~Ibrahim, M.~Parzen, and
  R.~Strawderman (2006).
\newblock Use of the probability integral transformation to fit nonlinear
  mixed-effects models with nonnormal random effects.
\newblock {\em Journal of Computational and Graphical Statistics\/}~{\em
  15\/}(1), 39--57.

\bibitem[\protect\citeauthoryear{{R Core Team}}{{R Core Team}}{2020}]{R}
{R Core Team} (2020).
\newblock {\em R: A Language and Environment for Statistical Computing}.
\newblock Vienna, Austria: R Foundation for Statistical Computing.

\bibitem[\protect\citeauthoryear{Smyth}{Smyth}{1998}]{statmod}
Smyth, G.~K. (1998).
\newblock Numerical integration.
\newblock In P.~Armitage and T.~Colton (Eds.), {\em Encyclopedia of
  Biostatistics}, pp.\  3088--3095. London: Wiley.

\end{thebibliography}

\newpage
\section*{Figures and table}

\begin{figure}[htb]
\centering
\makebox{\includegraphics[width=\textwidth]{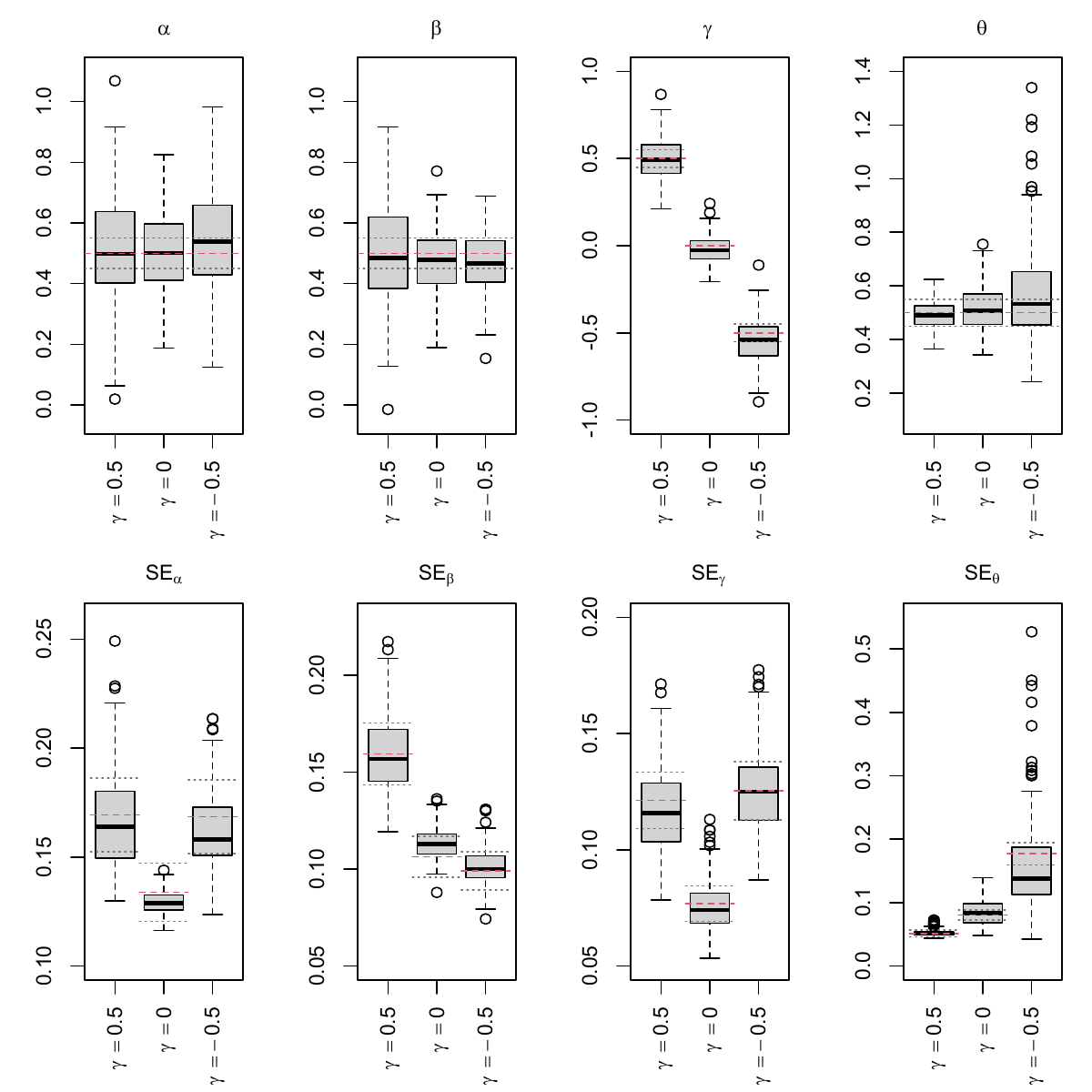}}
\caption{Box plots of the parameter estimates (top) and estimated standard errors (bottom) in the joint frailty model for positive ($\gamma=0.5$), no ($\gamma=0$), or negative ($\gamma=-0.5$) dependence under the base scenario. Left to right: covariate effect on mortality ($\alpha$) and on recurrences ($\beta$), dependence parameter ($\gamma$), and frailty variance ($\theta$) based on 200 truncated samples with a target size of $500$. The red dashed line marks the true parameter value (top) or empirical standard deviation (bottom); the gray dotted lines mark 10\% deviations from the respective value.}
\label{fig:simuEstim}
\end{figure}

\clearpage

\begin{table}[!hb]
\begin{center}
\caption{Parameter estimates (with standard errors) for the joint frailty model fitted to the cranberry data set, separately by risk group.}
\label{tab:cranb}
\begin{tabular}{llrr}
\hline
& & \multicolumn{1}{c}{High baseline UTI risk} & \multicolumn{1}{c}{Low baseline UTI risk}\\
\hline
\multicolumn{4}{l}{Recurrent UTIs}\\
& Treatment (cranberry) & $0.000$ $(0.161)$ & $0.189$ $(0.217)$\\
& Gender (male) & $0.061$ $(0.218)$ & $-0.384$ $(0.381)$\\
\multicolumn{4}{l}{Mortality}\\
& Treatment (cranberry) & $0.107$ $(0.152)$ & $-0.001$ $(0.197)$\\
& Gender (male) & $0.396$ $(0.178)$ & $0.787$ $(0.210)$\\
\multicolumn{4}{l}{Association}\\
& Dependence $\gamma$ & $0.181$ $(0.084)$ & 0.058 (0.044)\\
& Frailty variance $\theta$ & $0.380$ $(0.086)$ & $1.122$ $(0.316)$\\
\hline
\end{tabular}
\end{center}
\end{table}

\end{document}


\renewcommand{\theequation}{S.\arabic{equation}}
\renewcommand{\thefigure}{S.\arabic{figure}}
\renewcommand{\thetable}{S.\arabic{table}}
\renewcommand{\thesection}{S.\arabic{section}}

\begin{center}
\sffamily{\textbf{\Large{Supplementary Material for}\\
\LARGE{Incorporating delayed entry into the joint frailty model for recurrent events and a terminal event}}}
\end{center}
\vskip0.5cm
\renewcommand*{\thefootnote}{\fnsymbol{footnote}}
\noindent\normalfont
\large{\textbf{Marie B\"ohnstedt\footnote{Address for correspondence: \sffamily{e-mail: boehnstedt.marie@gmail.com}}$^{,1,2}$, Jutta Gampe$^1$, Monique A.~A.~Caljouw$^3$, and Hein Putter$^2$}}
\vskip0.25cm\normalsize\noindent
$^1$ Max Planck Institute for Demographic Research,  Rostock, Germany\\
$^2$ Department of Biomedical Data Sciences, Leiden University Medical Center, Leiden, The Netherlands\\
$^3$ Department of Public Health and Primary Care, Leiden University Medical Center, Leiden, The Netherlands
\vskip0.5cm
\noindent
\textit{Last revised: March 26, 2021}
\vskip0.75cm
\normalfont\normalsize

\section{Approximation of the marginal likelihood using Gaussian quadrature}

In this section, we will elaborate on the use of Gauss-Hermite quadrature for approximating the marginal likelihood of the joint frailty model. We will first recap the quadrature approach as proposed by \citet{LiuHuang:2008} in the setting with right-censoring only, and then show how to adapt the method to the setting with left truncation.

An approximation to marginal likelihoods integrated over normal random effects based on Gauss-Hermite quadrature was already presented in Section~3 of the main paper.
But the marginal likelihoods of the joint frailty model given in Section 2 of the main paper involve integrals over non-normal random effects. Thus, we use the probability integral transformation \citep[see][]{Nelsonetal:2006,LiuHuang:2008} to rewrite the integrals over the random effect~$u$ with distribution function~$G_{\theta}(u)$ as integrals over standard normal random effects. This relies on the fact that the $G_{\theta}(u)$ have a standard uniform distribution, such that their transformations $a=\Phi^{-1}[G_{\theta}(u)]$ follow a standard normal distribution, if $\Phi(\cdot)$ denotes the standard normal distribution function.

For the marginal likelihood contribution~(5) of individual~$i$ in the joint frailty model without truncation, the substitution $u=G^{-1}_{\theta}[\Phi(a)]$ yields
\[L_i=\int_0^{\infty} L_i^{(c)}(u)\,g_{\theta}(u)\,\mathrm{d}u = \int_{-\infty}^{\infty} L_i^{(c)}(G_{\theta}^{-1}[\Phi(a)])\,\phi(a)\,\mathrm{d}a.\]
We can then employ Gauss-Hermite quadrature as in~(11) to arrive at the approximate marginal likelihood contribution of the form
\[L_i \approx \sum_{q=1}^Q L_i^{(c)}(G_{\theta}^{-1}[\Phi(\tilde{x}_q)])\,\phi(\tilde{x}_q)\,\tilde{w}_q.\]

In the setting with left-truncated data, the marginal likelihood of the joint frailty model has a slightly more complex structure. In~(10), the marginal likelihood contribution~${}_V\!L_i$ of individual~$i$ is expressed as a ratio of two integrals over the density~$g_{\theta}(u)$. Hence, we will approximate the likelihood by applying the above approach separately to the two integrals,
\begin{align*}
{}_V\!L_i&=\frac{\int_0^{\infty} {}_V\!L_i^{(c)}(u)\,\exp{\{-H_i(v_i|u)\}}\,g_{\theta}(u)\,\mathrm{d}u}{\int_0^{\infty} \exp{\{-H_i(v_i|u)\}}\,g_{\theta}(u)\,\mathrm{d}u}\\
&\approx\frac{\sum_{q=1}^Q {}_V\!L_i^{(c)}(G_{\theta}^{-1}[\Phi(\tilde{x}_q)])\,\exp{\{-H_i(v_i|G_{\theta}^{-1}[\Phi(\tilde{x}_q)])\}}\,\phi(\tilde{x}_q)\,\tilde{w}_q}{\sum_{q=1}^Q \exp{\{-H_i(v_i|G_{\theta}^{-1}[\Phi(\tilde{x}_q)])\}}\,\phi(\tilde{x}_q)\,\tilde{w}_q},
\end{align*}
with ${}_V\!L_i^{(c)}(u)$ given in~(6). The approximate marginal likelihood of the joint frailty model with left truncation is then given by
\[\prod_{i=1}^{m_V} \frac{\sum_{q=1}^Q {}_V\!L_i^{(c)}(G_{\theta}^{-1}[\Phi(\tilde{x}_q)])\,\exp{\{-H_i(v_i|G_{\theta}^{-1}[\Phi(\tilde{x}_q)])\}}\,\phi(\tilde{x}_q)\,\tilde{w}_q}{\sum_{q=1}^Q \exp{\{-H_i(v_i|G_{\theta}^{-1}[\Phi(\tilde{x}_q)])\}}\,\phi(\tilde{x}_q)\,\tilde{w}_q}.\]
In the naive likelihood~(8), there is only one integral over the frailty distribution, and hence only one approximation is required,
\[{}_V\!L_i^{\text{naive}}=\int_0^{\infty} {}_V\!L_i^{(c)}(u)\,g_{\theta}(u)\,\mathrm{d}u\approx \sum_{q=1}^Q {}_V\!L_i^{(c)}(G_{\theta}^{-1}[\Phi(\tilde{x}_q)])\,\phi(\tilde{x}_q)\,\tilde{w}_q.\]

\section{Computational details}

We used \texttt{R} \citep{R} to implement the estimation procedure. The quadrature points and weights were calculated using function \texttt{gauss.quad()} from package \texttt{statmod} \citep{statmod}. For numerical optimization of the approximate marginal log-likelihood, we applied function \texttt{nlm()}, which performs minimization based on a Newton-type algorithm, to the negative log-likelihood.
The Hessian of the marginal log-likelihood was approximated numerically using function \texttt{hessian()} from package \texttt{numDeriv} \citep{numDeriv}.

As the frailty variance and the parameters of the piecewise constant baseline rates are restricted to be non-negative, the log-likelihood was maximized with respect to the log-transform of these parameters, which guaranteed non-negative estimates. The delta-method was then applied to derive the respective standard errors.

However, when specifying the baseline rates as piecewise constant functions, the numerically computed Hessian of the log-likelihood may in some cases not be invertible. To overcome this issue, one can add to the log-likelihood small, fixed ridge penalties (e.g., with penalty parameter $10^{-6}$) on the logarithm of the rate parameters and derive the Hessian matrix from this penalized log-likelihood.

\section{Supplement to the simulation studies}

\subsection*{Generation of different truncation patterns}

A brief description of how the truncation patterns~A and B were generated was already given in Section~4 of the main paper. For both patterns, truncation times were simulated from a truncated normal distribution defined on the corresponding age range $[75,95]$ or $[64,105]$. However, the parameters~$\mu$ and $\sigma^2$ of the underlying normal distribution were chosen such that the density of the resulting truncated normal distribution was either increasing over the whole age range with mode equal to the maximum age or unimodal with a mode within the age range. These two distinct shapes yielded the desired truncation patterns~A (TrA) or B (TrB) in the final samples.
Table~\ref{tab:pv} reports, for the different settings, the parameter values for the distribution of the truncation times as well as the initial sample size~$M$ needed to obtain truncated samples with an average size of $m_V=500$.

When implementing the simulation study in \texttt{R}, we used function \texttt{rtruncnorm()} from package \texttt{truncnorm} \citep{truncnorm} for drawing random numbers from a truncated normal distribution.

\begin{table}[htb]
\begin{center}
\caption{Initial sample size~$M$ and parameter values for the distribution of the truncation times.}
\label{tab:pv}
\begin{tabular}{lllrrrr}
\hline
\multicolumn{1}{c}{$\gamma$} & \multicolumn{1}{c}{age range} & \multicolumn{1}{c}{pattern} & \multicolumn{1}{c}{$M$} & \multicolumn{1}{c}{$\mu$} & \multicolumn{1}{c}{$\sigma^2$}\\
\hline
$\phantom{-}0.5$ & $[75,95]$ & TrA & $1.07\cdot 10^4$ & $109$ & $124$\\
$\phantom{-}0.5$ & $[75,95]$ & TrB & $1.20\cdot 10^4$ & $90$ & $18$\\
$\phantom{-}0.5$ & $[64,105]$ & TrA & $7.65\cdot 10^4$ & $120$ & $225$\\
$\phantom{-}0.5$ & $[64,105]$ & TrB & $4.75\cdot 10^4$ & $93$ & $52$\\
$\phantom{-}0$ & $[75,95]$ & TrA &$3.50\cdot 10^4$ & $115$ & $110$\\
$-0.5$ & $[75,95]$ & TrA & $3.44\cdot 10^4$ & $115$ & $109$\\
\hline
\end{tabular}
\end{center}
\end{table}

\subsection*{Additional figures for the simulation results}

\begin{figure}[htb]
\centering
\makebox{\includegraphics[width=\textwidth]{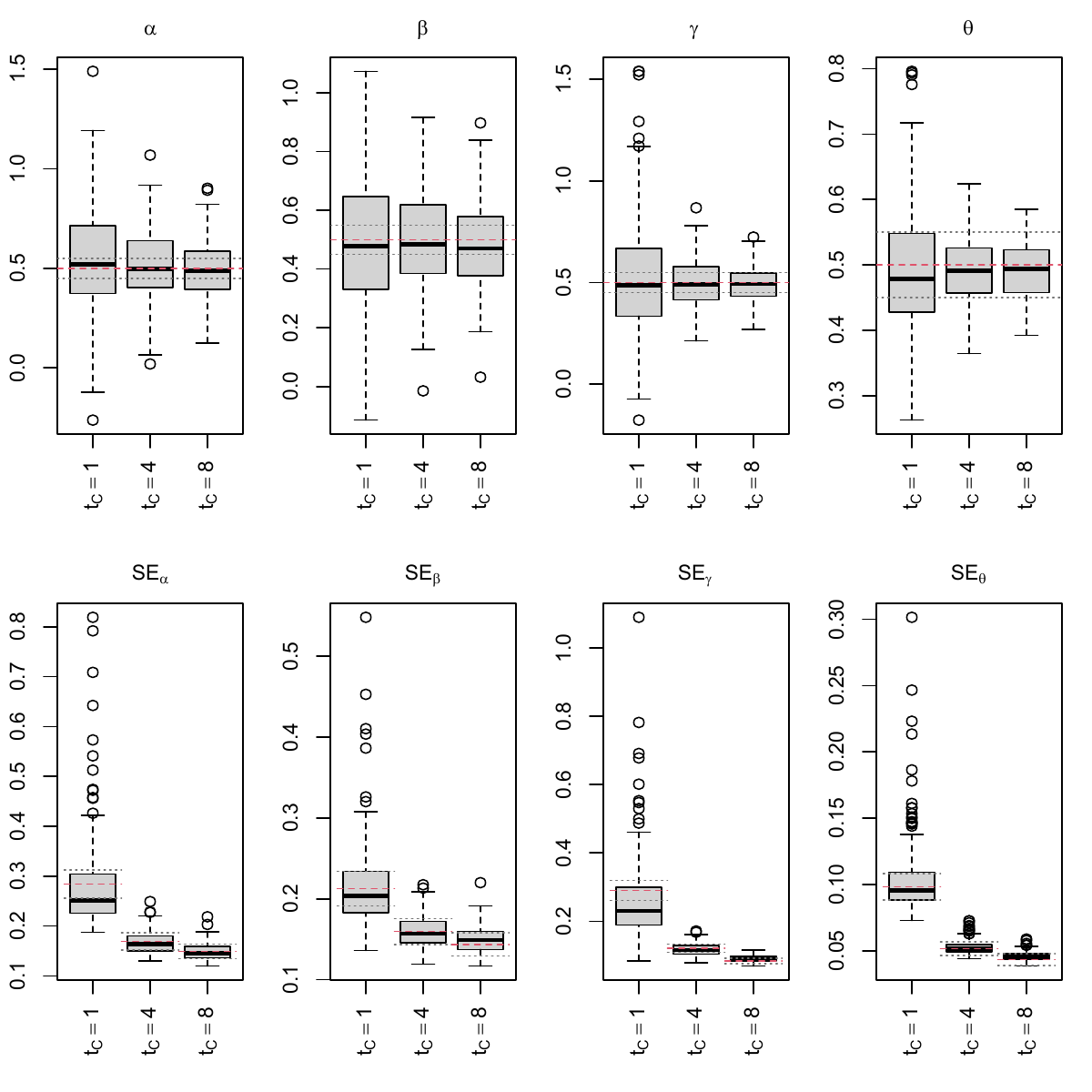}}
\caption{Box plots of the parameter estimates (top) and estimated standard errors (bottom) in the joint frailty model with positive dependence ($\gamma=0.5$) as in the base scenario with truncation pattern~A for ages $75+$, but varying the planned individual follow-up $t_C\in\{1,4,8\}$. Left to right: covariate effect on mortality ($\alpha$) and on recurrences ($\beta$), dependence parameter ($\gamma$), and frailty variance ($\theta$) based on 200 truncated samples with a target size of $500$. The red dashed line marks the true parameter value (top) or empirical standard deviation (bottom); the gray dotted lines mark 10\% deviations from the respective value.}
\label{fig:simuFup}
\end{figure}

\begin{figure}[htb]
\centering
\makebox{\includegraphics[width=\textwidth]{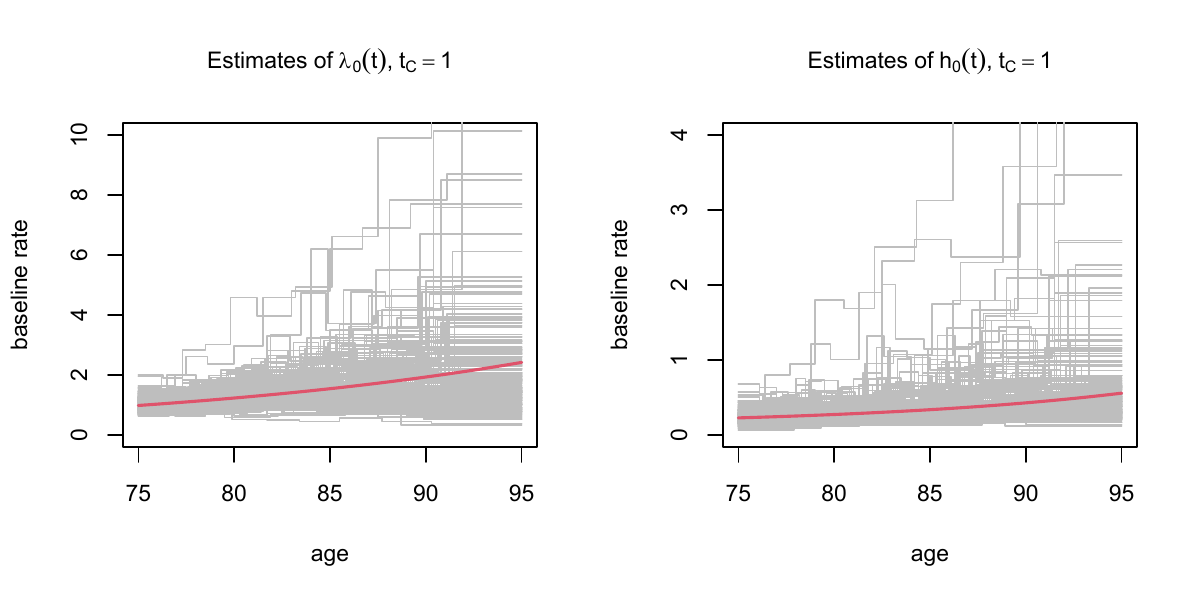}}
\makebox{\includegraphics[width=\textwidth]{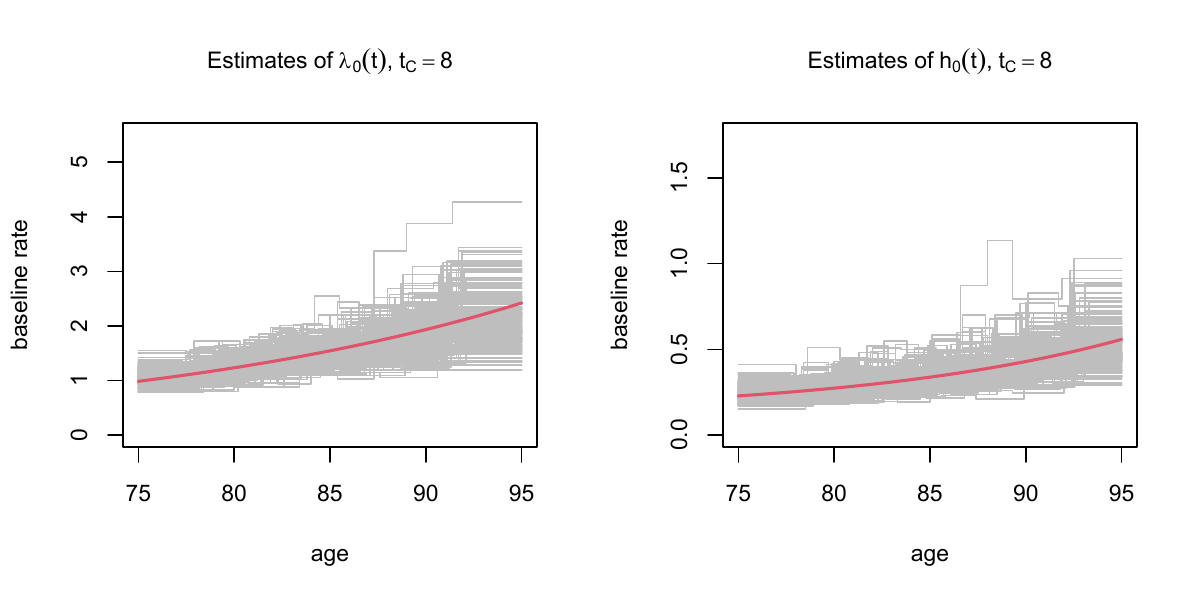}}
\caption{Estimates (gray) of the baseline rate of recurrence (left) and of death (right) based on 200 truncated samples with a target size of 500 generated from a joint frailty model with positive dependence ($\gamma=0.5$). As in the base scenario truncation follows pattern~A for ages 75+, but planned individual follow-up is $t_C=1$ (top) or $t_C=8$ (bottom) year(s). The red bold line gives the true baseline rate. (Note the different scales of the vertical axes.)}
\label{fig:simuRatesFup}
\end{figure}

\begin{figure}[htb]
\centering
\makebox{\includegraphics[width=\textwidth]{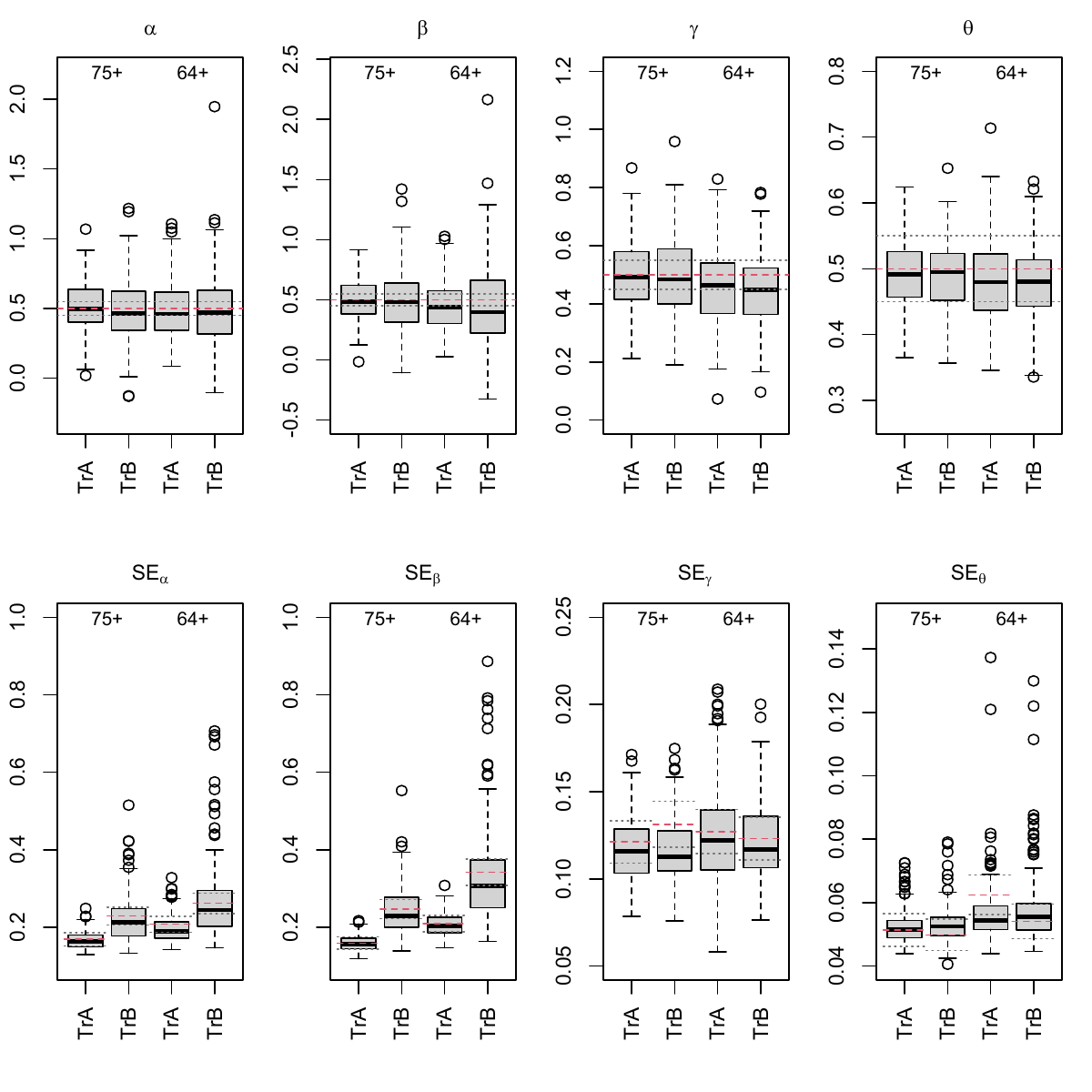}}
\caption{Box plots of the parameter estimates (top) and estimated standard errors (bottom) in the joint frailty model with positive dependence ($\gamma=0.5$) as in the base scenario with $t_C=4$, but truncation times distributed with different shapes (TrA: truncation pattern~A, TrB: truncation pattern~B) and across different age ranges (75+: $[75,95]$, 64+: $[64,105]$). Left to right: covariate effect on mortality ($\alpha$) and on recurrences ($\beta$), dependence parameter ($\gamma$), and frailty variance ($\theta$) based on 200 truncated samples with a target size of $500$. The red dashed line marks the true parameter value (top) or empirical standard deviation (bottom); the gray dotted lines mark 10\% deviations from the respective value.}
\label{fig:simuTrunc}
\end{figure}

\begin{figure}[htb]
\centering
\makebox{\includegraphics[width=\textwidth]{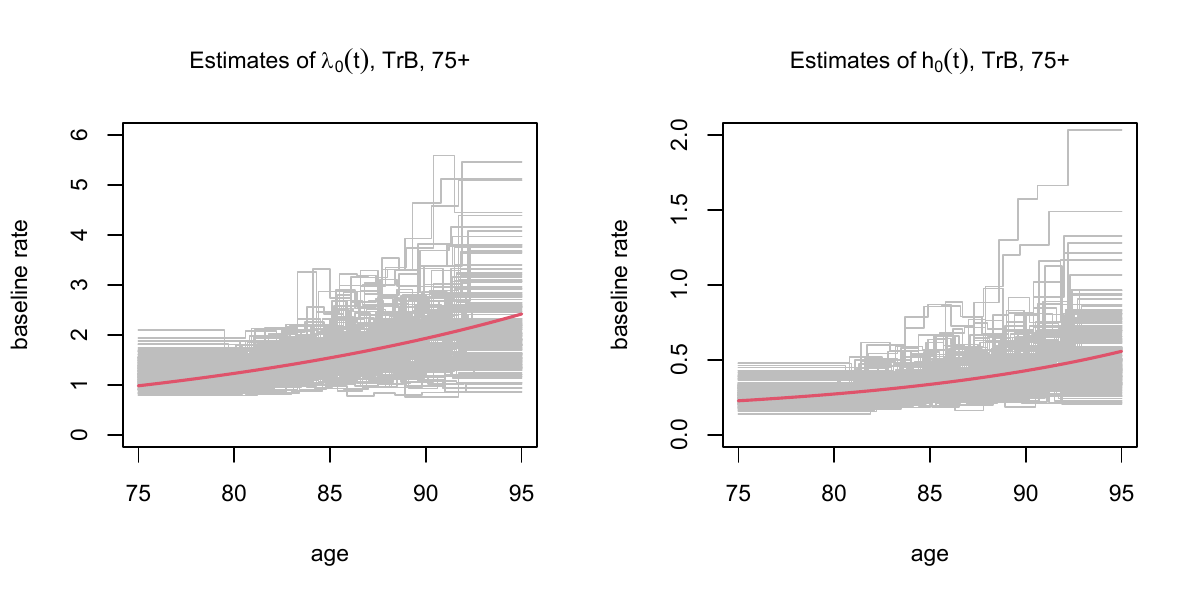}}
\makebox{\includegraphics[width=\textwidth]{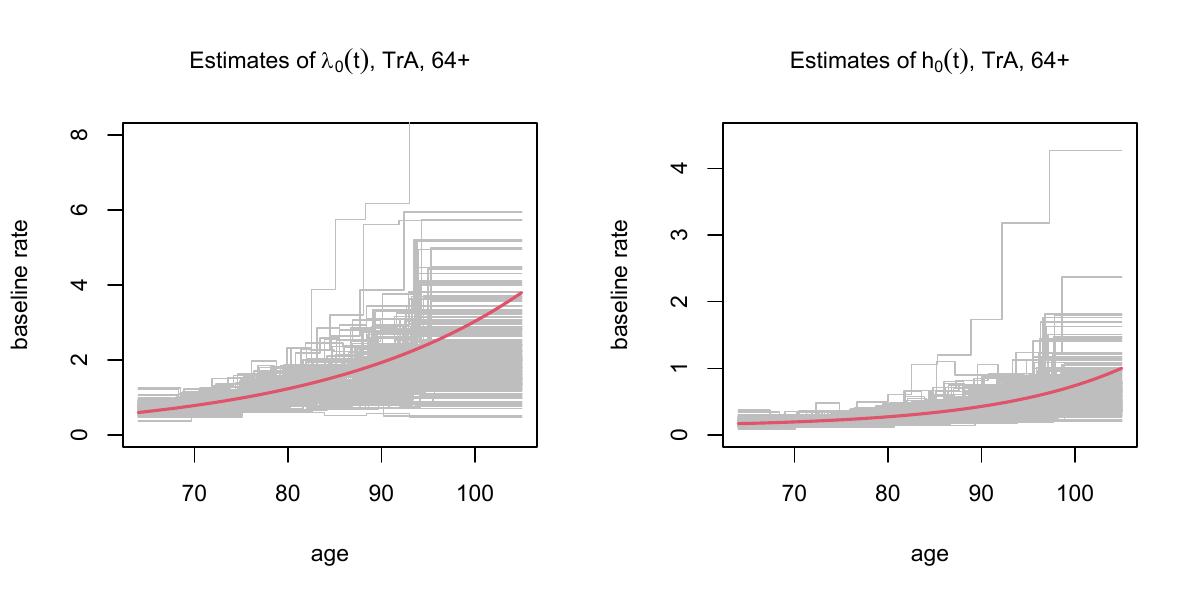}}
\caption{Estimates (gray) of the baseline rate of recurrence (left) and of death (right) based on 200 truncated samples with a target size of 500 generated from a joint frailty model with positive dependence ($\gamma=0.5$) and planned individual follow-up of $t_C=4$ years as in the base scenario. Truncation times are distributed according to pattern~B across ages $[75,95]$ (top; TrB, 75+) or according to pattern~A across ages $[64,105]$ (bottom; TrA, 64+). The red bold line gives the true baseline rate. (Note the different scales of the horizontal and vertical axes.)}
\label{fig:simuRatesTrunc}
\end{figure}

The following figures illustrate the impact of different censoring and truncation patterns on the performance of the estimation procedure. All settings are modifications of the base scenario with positive dependence~$\gamma=0.5$, that was presented in Section~4 of the main paper. In particular, samples with a target size of $m_V=500$ were generated from a joint frailty model with covariate effects $\alpha=\beta=0.5$, frailty variance $\theta=0.5$, and Gompertz-Makeham baseline rates.
\begin{itemize}[leftmargin=0.5cm]
\item In Figures~\ref{fig:simuFup} and \ref{fig:simuRatesFup}, the effect of different censoring mechanisms is studied by comparing the estimation results for different lengths of the planned individual follow-up~$t_C\in\{1,4,8\}$ years.
\item Changes in the distribution of truncation times are examined in Figures~\ref{fig:simuTrunc} and \ref{fig:simuRatesTrunc}. The displayed settings assume different distributions of the study entry times in the final sample, both in terms of the shape (TrA: truncation pattern~A, or TrB: truncation pattern~B) and the support of the distribution (75+: ages 75 to 95, or 64+: ages 64 to 105).
\end{itemize}

\vfill
\bibliographystyle{chicago}
\bibliography{JFMwLTbibSuppl}